\pdfoutput = 1

\documentclass[12pt,a4paper]{article}

\usepackage{lscape}
\usepackage{rotating}
\usepackage{float}
\usepackage{amsmath}
\usepackage{amssymb}
\usepackage{cite}
\usepackage{hyperref}
\usepackage{graphicx}
\usepackage{booktabs}
\usepackage[caption=false]{subfig}

\newcommand{\lsim}{\mathrel{\hbox{\rlap{\hbox{\lower4pt\hbox{$\sim$}}}\hbox{$<$}}}}

\usepackage{xspace}

\newcommand{\Bbar}{\kern 0.18em\overline{\kern -0.18em B}{}\xspace}

\newcommand{\Kbar}{\kern 0.18em\overline{\kern -0.18em K}{}\xspace}

\makeatletter

\newcommand{\Rmnum}[1]{\expandafter\@slowromancap\romannumeral #1@}
\makeatother

\begin{document}
\begin{titlepage}

\vspace*{-1.3truecm}

\begin{flushright}
Nikhef-2017-023
\end{flushright}

\vspace{0.9truecm}

\begin{center}
\boldmath
{\Large{\bf In Pursuit of New Physics with $B^0_{s,d}\to\ell^+\ell^-$}
}
\unboldmath
\end{center}

\vspace{1.8truecm}

\begin{center}
{\bf Robert Fleischer\,${}^{a,b}$, Ruben Jaarsma\,${}^{a}$ and Gilberto~Tetlalmatzi-Xolocotzi\,${}^{a}$}

\vspace{0.5truecm}

${}^a${\sl Nikhef, Science Park 105, NL-1098 XG Amsterdam, Netherlands}

${}^b${\sl  Department of Physics and Astronomy, Vrije Universiteit Amsterdam,\\
NL-1081 HV Amsterdam, Netherlands}

\end{center}

\vspace{1.7cm}


\begin{abstract}
\noindent
Leptonic rare decays of $B^0_{s,d}$ mesons offer a powerful tool to search for physics beyond the Standard 
Model. The $B^0_{s}\to\mu^+\mu^-$ decay has been observed at the Large Hadron Collider and the first 
measurement of the effective lifetime of this channel was presented, in accordance with the Standard Model. 
On the other hand, $B^0_{s}\to\tau^+\tau^-$ and $B^0_{s}\to e^+e^-$ have received considerably less 
attention: while LHCb has recently reported a first upper limit of $6.8\times10^{-3}$ (95\% C.L.) for the 
$B^0_s\to\tau^+\tau^-$ branching ratio, the upper bound $2.8\times 10^{-7}$ (90\% C.L.) for the 
branching ratio of $B^0_s\to e^+e^-$ was reported by CDF back in 2009. We discuss the current status 
of the interpretation of the measurement of $B^0_{s}\to\mu^+\mu^-$, and explore the space for 
New-Physics effects in the other $B^0_{s,d}\to\ell^+\ell^-$ decays in a scenario assuming
flavour-universal Wilson coefficients of the relevant four-fermion operators. While the New-Physics effects 
are then strongly suppressed by the ratio $m_\mu/m_\tau$ of the lepton masses in $B^0_s\to\tau^+\tau^-$, 
they are hugely enhanced by $m_\mu/m_e$ in $B^0_s\to e^+e^-$ and may result in a 
$B^0_s\to e^+e^-$ branching ratio as large as about 5 times
the one of $B^0_{s}\to\mu^+\mu^-$, which is about a factor of 20 below the CDF bound; 
a similar feature arises in $B^0_{d}\to e^+e^-$.  Consequently, it would be most interesting to search 
for the $B^0_{s,d}\to e^+e^-$ channels at the LHC and Belle II, which may result in an unambiguous 
signal for physics beyond the Standard Model.
\end{abstract}


\vspace*{2.1truecm}

\vfill

\noindent
March 2017

\end{titlepage}

\thispagestyle{empty}
\vbox{}
\newpage

\setcounter{page}{1}

\section{Introduction}

The decay $B^0_s\to\mu^+\mu^-$ belongs to the most interesting processes for testing the
flavour sector of the Standard Model (SM). In this framework, $B^0_s\to\mu^+\mu^-$ 
emerges from quantum loop effects -- penguin and box topologies -- and is helicity suppressed. 
Consequently, this channel is strongly suppressed, and in the SM only about three out of one 
billion $B^0_s$ mesons
decay into the $\mu^+\mu^-$ final state. Another key feature of the $B^0_s\to\mu^+\mu^-$ decay
is that the binding of the anti-bottom quark and the strange quark in the $B^0_s$ meson is described by
a single non-perturbative parameter, the $B^0_s$-meson decay constant $f_{B_s}$ \cite{Bsmumu-SM}.
In scenarios of New Physics (NP), $B^0_s\to\mu^+\mu^-$ may be affected by new particles entering 
the loop topologies or may even arise at the tree level. 

For decades, experiments have searched for the $B^0_s\to\mu^+\mu^-$ decay \cite{rev}. 
It has been a highlight of run 1 of the Large Hadron Collider (LHC) that the $B^0_s\to\mu^+\mu^-$ 
mode could eventually be observed in a combined analysis by the CMS and LHCb collaborations 
\cite{CMS-LHCb}. The corresponding experimental branching ratio is consistent with the SM prediction. 

In addition to the branching ratio, $B^0_s\to\mu^+\mu^-$ offers another observable \cite{Bsmumu-ADG}. 
It is accessible thanks to the sizeable difference $\Delta\Gamma_s$ between the decay widths of the 
$B_s$ mass eigenstates and is encoded in the effective $B^0_s\to\mu^+\mu^-$ lifetime. 
The LHCb collaboration has very recently reported a pioneering measurement of this observable, 
as well as the observation of $B^0_s\to\mu^+\mu^-$ in the analysis of their new data set collected at
the ongoing run 2 of the LHC \cite{LHCb-2017}.

Furthermore, there is a CP-violating rate asymmetry which is generated through the interference between
$B^0_s$--$\bar B^0_s$ mixing and decay processes \cite{Bsmumu-ADG,BFGK}. It would be very 
interesting to measure this observable. However, such an analysis requires tagging information 
on the decaying $B_s$ meson, thereby making it more challenging than the untagged effective 
lifetime measurement.

In the SM, the key difference of $B^0_s\to\mu^+\mu^-$ with respect to $B^0_{s}\to\tau^+\tau^-$ and 
$B^0_{s}\to e^+e^-$ is due to the different lepton masses. In the case of the former decay, the large
$\tau$ mass effectively lifts the helicity suppression, while it gets much stronger for the latter process 
due to the small electron mass. Consequently, we have SM branching ratios at the $10^{-6}$ 
and $10^{-13}$ level, respectively, while the corresponding $B^0_s\to\mu^+\mu^-$ branching ratio 
takes a value at the $10^{-9}$ level \cite{Bsmumu-SM}.

\begin{figure}
\centering
\includegraphics[height=4.5cm]{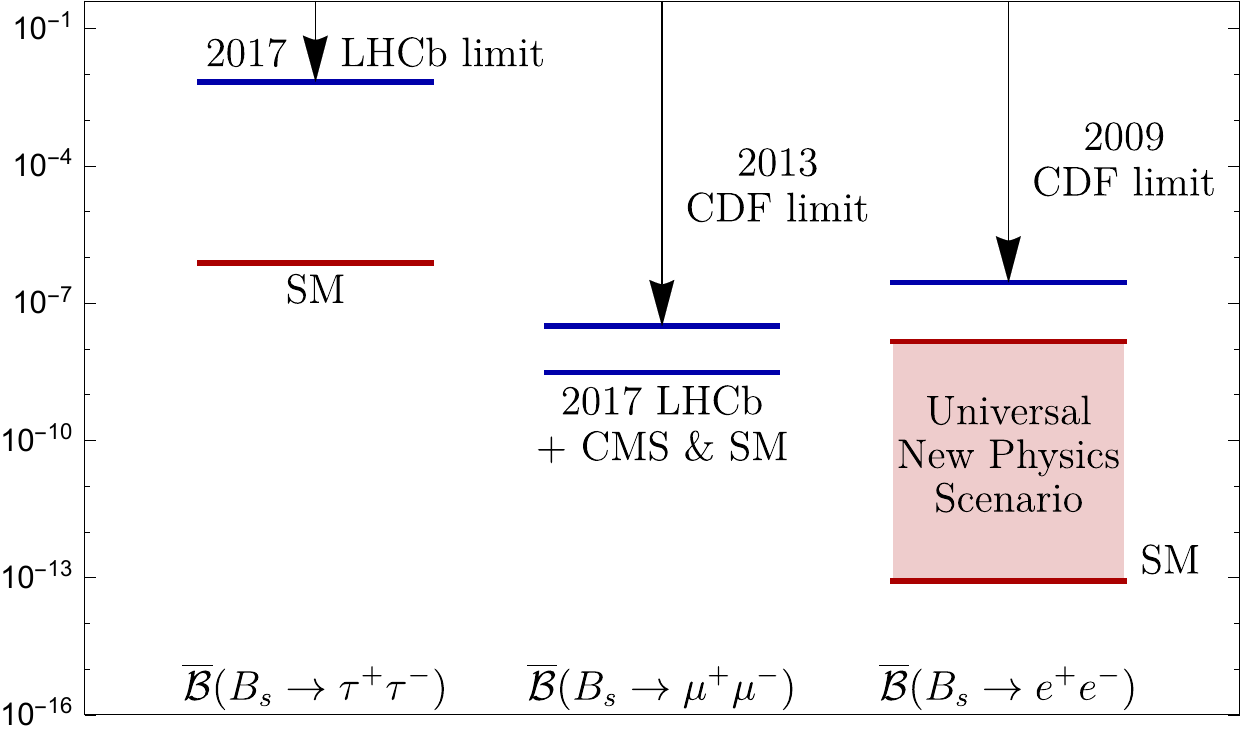}
\includegraphics[height=4.5cm]{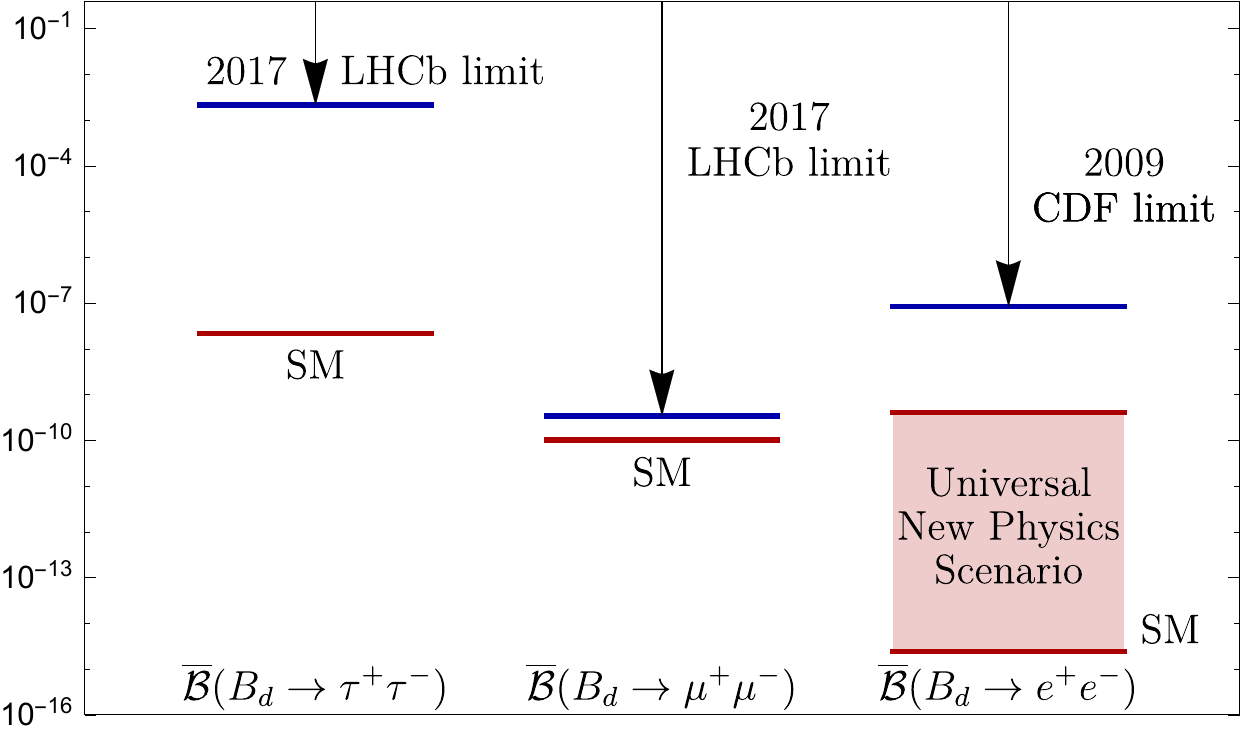}
\caption{Illustration of the $B^0_s\to\ell^+\ell^-$ (left panel) and $B^0_d\to\ell^+\ell^-$ (right panel)
branching ratios: current experimental status, SM predictions and the possible enhancement within
the general flavour-universal NP scenario discussed in the text.}
\label{fig:BR-status}
\end{figure}

From the experimental point of view, the analysis of  $B^0_{s}\to\tau^+\tau^-$ is challenging because of the
reconstruction of the $\tau$ leptons. Nevertheless, LHCb has recently presented the first upper
bound for this channel of $6.8\times10^{-3}$ (95\% C.L.) \cite{LHCb-Btautau}. The
$B^0_{s}\to e^+e^-$ decay has received surprisingly little attention, both from the experimental 
and theoretical communities, and has so far essentially not played any role in the exploration of 
flavour physics.  The most recent upper bound on the $B^0_s\to e^+e^-$ branching ratio of 
$2.8\times 10^{-7}$ (90\% C.L.) was obtained by the CDF collaboration back in 2009 \cite{CDF-Bsee}.
We have illustrated this situation in Fig.~\ref{fig:BR-status}.

We will have a fresh look at the search for NP effects with the $B^0_s\to\mu^+\mu^-$ and
$B^0_d\to\mu^+\mu^-$ channels in view of the
new LHCb data \cite{LHCb-2017}, complementing the recent study by Altmannshofer, Niehoff and 
Straub \cite{ANS}. However, the main focus of our discussion will be on the 
$B^0_{s}\to\tau^+\tau^-$ and $B^0_{s}\to e^+e^-$ decays (as well as their $B^0_d$ counterparts), 
which were not considered in Ref.~\cite{ANS}. The utility of the decays with $\tau^+\tau^-$ and
$e^+e^-$ in the final states for probing NP effects was addressed in the literature before, for instance, 
in Refs.~\cite{GLN,CHYY} and \cite{BHP,AGMC}, respectively. The current key question is how much 
space for NP effects is left in these channels by the currently available data, in particular for the 
experimentally established $B^0_s\to \mu^+\mu^-$ mode.

In order to explore this topic, which is in general very complex, and to illustrate possible NP effects,
we consider a framework where the Wilson coefficients of the relevant four-fermion operators are flavour
universal, i.e.\ do neither depend on the flavour of the decaying $B^0_s$ or $B^0_d$ mesons nor on 
the final-state leptons. We find that the corresponding NP effects are strongly suppressed 
in $B^0_{s}\to\tau^+\tau^-$ in this scenario. However, as the helicity suppression is lifted by new 
(pseudo)-scalar contributions, we may get a huge enhancement of the branching ratio of 
$B^0_{s}\to e^+e^-$ in this scenario, while still having the branching ratio of $B^0_{s}\to \mu^+\mu^-$ 
within the current experimental range. In particular, the branching ratio of 
$B^0_s\to e^+e^-$ may be enhanced to about 5 times the $B^0_s\to \mu^+\mu^-$ 
branching ratio, which is a factor of 20 below the CDF limit from 2009. Consequently, 
it would be most interesting to have a dedicated search for $B^0_s\to e^+e^-$ and $B^0_d\to e^+e^-$, 
fully exploiting the physics potential of the LHC, where these decays will be interesting for 
ATLAS, CMS and LHCb, and the future Belle II experiment at KEK. In view of the theoretical 
cleanliness of these decays and the possible spectacular enhancement with respect to the SM, we may 
get an unambiguous signal for New Physics.

\begin{figure}
\centering
\hspace*{-0.5truecm}\includegraphics[height=4.1cm]{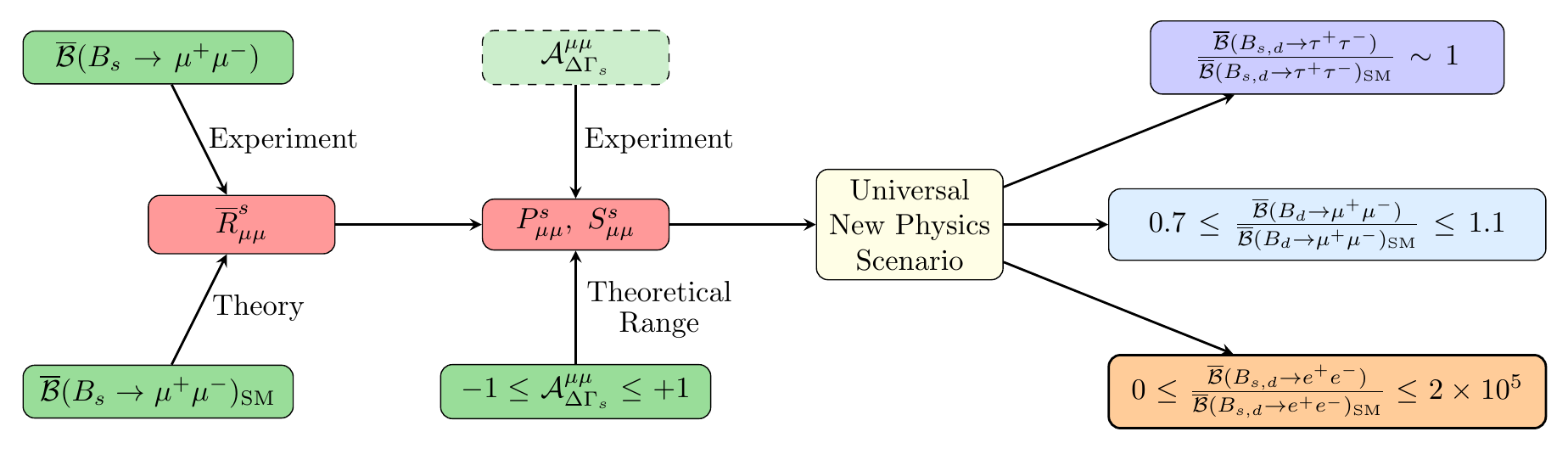}
\caption{Flowchart illustrating the analysis and interplay of the various $B^0_{s,d}\to\ell^+\ell^-$
observables within the considered flavour universal NP scenario, as discussed in the text.}\label{fig:chart}
\end{figure}

In Fig.~\ref{fig:chart}, we have illustrated our NP analysis. The measured branching ratio of 
the $B^0_s\to\mu^+\mu^-$ channel allows us to constrain the corresponding short-distance 
functions, which are then converted into their counterparts for the $B^0_{s,d}\to\tau^+\tau^-$ 
and $B^0_{s,d}\to e^+e^-$ channels, having very different implications. The flowchart in 
Fig.~\ref{fig:chart} serves as a guideline for the following discussion.

The outline of this paper is as follows: we discuss the theoretical framework for our studies
in Section~\ref{sec:theo}. In Section~\ref{sec:Bsmumu}, we have a closer look at the state-of-the-art
picture following from the experimental results for the $B^0_{s,d}\to\mu^+\mu^-$ decays, while 
turning to the $B^0_{s,d}\to\tau^+\tau^-$ and $B^0_{s,d}\to e^+e^-$ modes in Sections~\ref{sec:Bstautau} 
and \ref{sec:Bsee}, respectively. Finally, we summarize our conclusions in Section~\ref{sec:concl}.

\boldmath
\section{Theoretical Framework}\label{sec:theo}
\unboldmath
\subsection{Low-Energy Effective Hamiltonian}
Leptonic rare decays of $\bar B^0_q$ mesons ($q=d,s$) are described by the following low-energy 
effective Hamiltonian \cite{Bsmumu-SM,Bsmumu-ADG,ANS}:
\begin{equation}\label{Heff}
{\cal H}_{\rm eff}=-\frac{G_{\rm F}}{\sqrt{2}\pi} V_{tq}^\ast V_{tb} \alpha
\bigl[C^{q,\ell\ell}_{10} O_{10} + C^{q,\ell\ell}_{S} O_S + C^{q,\ell\ell}_P O_P+ 
C_{10}^{q,\ell\ell'} O_{10}' + C_{S}^{q,\ell\ell'} O_S' + C_P^{q,\ell\ell'} O_P' \bigr],
\end{equation}
where $G_{\rm F}$ denotes the Fermi constant, $V_{qq'}$ are elements of the 
Cabibbo--Kobayashi--Maskawa (CKM) matrix, and $\alpha$ is the QED fine structure constant.
The heavy degrees of freedom have been integrated out and are described by the Wilson coefficients 
$C_{10}^{q,\ell\ell(')}$, $C_{P}^{q,\ell\ell(')}$ and $C_{S}^{q,\ell\ell(')}$, which may depend both
on the flavour of the quark $q$ and on the flavour of the final-state leptons $\ell^+\ell^-$. However, in 
the SM and NP scenarios with ``Minimal Flavour Violation" (MFV) \cite{MFV}, the short-distance 
functions are flavour universal.

The Wilson coefficients are associated with the four-fermion operators
\begin{equation}
\begin{array}{rclcrcl}
O_{10}&=&(\bar q \gamma_\mu P_L b) (\bar\ell\gamma^\mu \gamma_5\ell), & \mbox{}\qquad &
 O_{10}'&=&(\bar q \gamma_\mu P_R b) (\bar\ell\gamma^\mu \gamma_5\ell),  \\
O_S&=&m_b (\bar q P_R b)(\bar \ell \ell), & \mbox{}\qquad &  O_S'&=&m_b (\bar q P_L b)(\bar \ell \ell),\\
O_P&=&m_b (\bar q P_R b)(\bar \ell \gamma_5 \ell), & \mbox{}\qquad  
& O_P'&=&m_b (\bar q P_L b)(\bar \ell \gamma_5 \ell),
\end{array}
\end{equation}
where $m_b$ denotes the $b$-quark mass and 
\begin{equation}
P_{L}\equiv\frac{1}{2}\left(1-\gamma_5\right), \quad P_{R}\equiv\frac{1}{2}\left(1+\gamma_5\right).
\end{equation}
In the general Hamiltonian in Eq.~(\ref{Heff}), we have only kept operators which 
give non-vanishing contributions to $\bar B^0_q\to \ell^+\ell^-$ decays. 
In the SM, only the $O_{10}$ operator is present with a real coefficient $C_{10}^{\rm SM}$.

Concerning the impact of NP, the outstanding feature of the $\bar B^0_q\to \ell^+\ell^-$ channels is 
their sensitivity to (pseudo)-scalar lepton densities entering the operators $O_{(P)S}$ and 
$O_{(P)S}'$, which have still largely unconstrained Wilson coefficients, thereby offering an interesting avenue 
for NP effects to enter. The $\bar B^0_q\to\ell^+\ell^-$ decay amplitude has the following
structure \cite{Bsmumu-SM}:
\begin{equation}
A(\bar B^0_q\to\ell^+\ell^-)\propto V_{tq}^\ast V_{tb} f_{B_q} M_{B_q}   
m_\mu C_{10}^{\rm SM} \left [\eta_\lambda P^q_{\ell\ell} +  S^q_{\ell\ell}   \right],
\end{equation}
where $\lambda={\rm L, R}$ describes the helicity of the final-state leptons with $\eta_{\rm L}=+1$ and  
$\eta_{\rm R}=-1$. The quantities
\begin{equation}\label{P-expr}
P^q_{\ell\ell}\equiv \frac{C_{10}^{q,\ell\ell}-C_{10}^{q,\ell\ell'}}{C_{10}^{\rm SM}}+\frac{M_{B_q}^2}{2 m_\ell}
\left(\frac{m_b}{m_b+m_q}\right)\left[\frac{C^{q,\ell\ell}_P-C_P^{q,\ell\ell'}}{C_{10}^{\rm SM}}\right]
\end{equation}
\begin{equation}\label{S-expr}
S^q_{\ell\ell}\equiv \sqrt{1-4\frac{m_\ell^2}{M_{B_q}^2}}
\frac{M_{B_q} ^2}{2 m_\ell}\left(\frac{m_b}{m_b+m_q}\right)
\left[\frac{C_S^{q,\ell\ell}-C_S^{q,\ell\ell'}}{C_{10}^{\rm SM}}\right],
\end{equation}
where $M_{B_q}$ and $m_\ell$ are the $\bar B^0_q$ and $\ell$ masses, respectively, will play a 
key role in the following discussion. In general, the coefficients 
$P^q_{\ell\ell}\equiv |P^q_{\ell\ell}|e^{i\varphi_{P_q}^{\ell\ell}}$ and 
$S^q_{\ell\ell}\equiv |S^q_{\ell\ell}|e^{i\varphi_{S_q}^{\ell\ell}}$ have 
CP-violating phases $\varphi_{P_q}^{\ell\ell}$ and $\varphi_{S_q}^{\ell\ell}$. 
In the SM, we obtain the simple relations 
\begin{equation}
P^q_{\ell\ell}|_{\rm SM}=1, \quad S^q_{\ell\ell}|_{\rm SM}=0.
\end{equation}

\subsection{Decay Observables}
The $B^0_s$ and $\bar B^0_s$ mesons show the phenomenon of $B^0_s$--$\bar B^0_s$ mixing,
which leads to time-dependent decay rates. Experiments actually 
measure the following time-integrated branching ratio \cite{DFN}:
\begin{equation}
	\overline{\mathcal{B}}(B_s\to\ell^+\ell^-) \equiv \frac{1}{2}\int_0^\infty \langle
	\Gamma(B_s(t)\to \ell^+\ell^-)
	\rangle\,dt.\label{BRexp}
\end{equation}
Here the time-dependent untagged rate, where no distinction is made between initially, i.e.\ at time
$t=0$, present $B^0_s$ or $\bar B^0_s$ mesons, takes the following form \cite{Bsmumu-ADG,ANS,BFGK}:
\begin{displaymath}
	\hspace*{-3.5truecm}\langle\Gamma(B_s(t)\to \ell^+ \ell^-)\rangle 
	\equiv \Gamma(B_s^0(t)\to \ell^+ \ell^-) + \Gamma(\bar B_s^0(t)\to \ell^+ \ell^-)
\end{displaymath}
\begin{equation}\label{untaggedRate}
	= \frac{G_{\rm F}^2\alpha^2}{16\pi^3}\left|V_{ts} V_{tb}^*\right|^2f_{B_s}^2M_{B_s} m_\ell^2
	\sqrt{1- 4\frac{m^2_\ell}{M_{B_s}^2}}\left|C_{10}^{\rm SM}\right|^2
	\end{equation}
\begin{displaymath}
\hspace*{3.0truecm}\times\,\left(|P^s_{\ell\ell}|^2 + |S^s_{\ell\ell}|^2\right)e^{-t/\tau_{B_s}}
\left[\cosh\left(y_s\, t/\tau_{B_s}\right) + {\cal A}^{\ell\ell}_{\Delta\Gamma_s}
\sinh\left(y_s\, t/\tau_{B_s}\right)\right],
\end{displaymath}
where the decay width difference $\Delta\Gamma_s$ enters through the parameter \cite{Amh16}
\begin{equation}
y_s\equiv \frac{\Delta\Gamma_s}{2\Gamma_s}= 0.0645\pm0.0045,
\end{equation}
with $\tau_{B_s}=1/\Gamma_s$ denoting the $B^0_s$ lifetime. Using the quantities introduced
above, the observable ${\cal A}^{\ell\ell}_{\Delta\Gamma_s}$ is given as follows \cite{Bsmumu-ADG,BFGK}:
\begin{equation}\label{ADG}
{\cal A}^{\ell\ell}_{\Delta\Gamma_s} = \frac{|P^s_{\ell\ell}|^2\cos(2\varphi_{P_s}^{\ell\ell}-\phi_s^{\rm NP}) - 
|S^s_{\ell\ell}|^2\cos(2\varphi_{S_s}^{\ell\ell}-\phi_s^{\rm NP})}{|P^s_{\ell\ell}|^2 + |S^s_{\ell\ell}|^2}.
\end{equation}
Since it is challenging to determine the helicity of the final-state leptons experimentally, 
the rates in (\ref{untaggedRate}) are actually helicity-averaged. The observable 
${\cal A}^{\ell\ell}_{\Delta\Gamma_s}$ takes the SM value
\begin{equation}\label{eq:ADGSM}
{\cal A}^{\ell\ell}_{\Delta\Gamma_s}|_{\rm SM}=+1,
\end{equation}
but is essentially unconstrained when allowing for NP effects \cite{Bsmumu-ADG,BFGK,ANS}.

In view of the sizeable $y_s$, we have to properly distinguish between the time-integrated branching 
ratio $\overline{\mathcal{B}}(B_s\to\ell^+\ell^-)$ measured at experiments and the ``theoretical" branching
ratio $\mathcal{B}(B_s\to\ell^+\ell^-)_{\rm theo}$, which corresponds to the decay time $t=0$. These
two branching ratios can be converted into each other through the following relation \cite{BR-paper}:
\begin{equation}
	\mathcal{B}(B_s \to \ell^+\ell^-)_{\rm theo}= 
	\left[\frac{1-y_s^2}{1 + {\cal A}^{\ell\ell}_{\Delta\Gamma_s}\, y_s}\right] 
	\overline{\mathcal{B}}(B_s\to\ell^+\ell^-).
\end{equation}

The physics information encoded in the effective lifetime 
\begin{equation}
	\tau^s_{\ell\ell} \equiv \frac{\int^\infty_0 t\,
	\langle\Gamma(B_s(t)\to \ell^+ \ell^-)\rangle\, dt}{\int_0^\infty \langle
	\Gamma(B_s(t)\to \ell^+ \ell^-)\rangle\, dt}
\end{equation}
is equivalent to the observable ${\cal A}^{\ell\ell}_{\Delta\Gamma_s}$ \cite{Bsmumu-ADG}, 
which can be determined with the help of 
\begin{equation} \label{eq:aDGLifetime}
 {\cal A}^{\ell\ell}_{\Delta\Gamma_s}  = \frac{1}{y_s}\left[\frac{(1-y_s^2)\tau^s_{\ell\ell}-(1+
 y_s^2)\tau_{B_s}}{2\tau_{B_s}-(1-y_s^2)\tau^s_{\ell\ell}}\right].
\end{equation} 
Moreover, $\tau^s_{\ell\ell}$ allows us to convert the time-integrated branching ratio determined
at experiments into the ``theoretical" branching ratio with the help of the relation
\begin{equation}\label{BR-correct}
\mathcal{B}(B_s \to \ell^+\ell^-)_{\rm theo}=\left[2 - 
	\left(1-y_s^2\right)\frac{\tau^s_{\ell\ell}}{\tau_{B_s}}\right]\overline{\mathcal{B}}(B_s\to\ell^+\ell^-),
\end{equation}
where all quantities on the right-hand side can be measured \cite{Bsmumu-ADG,BR-paper}. In the
case of $\bar B^0_d\to \ell^+\ell^-$ decays, $y_d$ takes a value at the $10^{-3}$ level. Consequently, 
the corresponding observable ${\cal A}^{\ell\ell}_{\Delta\Gamma_{d}}$  is experimentally not accessible 
in the foreseeable future. 

In addition to these untagged observables, there are also CP-violating asymmetries which would
be very interesting to measure, providing insights into possible new sources for CP violation encoded
in the Wilson coefficients \cite{Bsmumu-ADG,BFGK}. The experimental analysis of these observables
would require tagging information, thereby making it more challenging than the exploration of 
${\cal A}^{\ell\ell}_{\Delta\Gamma_{s}}$. However, it would nevertheless be very interesting to make 
efforts in the super-high-precision era of $B$ physics to get also a handle on these quantities.

In order to search for NP effects by means of the branching ratio of the $B^0_s\to\ell^+\ell^-$ decays, 
it is useful to introduce the following ratio \cite{Bsmumu-ADG,BFGK}:
\begin{equation}\label{Rellells}
 \overline{R}_{\ell\ell}^s \equiv \frac{\overline{\mathcal{B}}(B_s\to\ell^+\ell^-)}{\overline{\mathcal{B}}
   (B_s\to\ell^+\ell^-)_{\rm SM}},
\end{equation}
which takes by definition the SM value 
\begin{equation}
\overline{R}^s_{\ell\ell} |_{\rm SM}=1.
\end{equation}
Using the expressions given above yields
\begin{displaymath}\label{eq:Rll} 
   \hspace*{-0.9truecm} 
   \overline{R}_{\ell\ell}^s = \left[\frac{1+{\cal A}^{\ell\ell}_{\Delta\Gamma_s}\,y_s}{1+y_s} \right] 
	 (|P^s_{\ell\ell}|^2 + |S^s_{\ell\ell}|^2)
\end{displaymath}
\begin{equation} \label{eq:RBarll}
= \left[\frac{1+y_s\cos(2\varphi^{\ell\ell}_{P_s}-\phi_s^{\rm NP})}{1+y_s} \right] |P_{\ell\ell}|^2 + 
\left[\frac{1-y_s\cos(2\varphi^{\ell\ell}_{S_s}-\phi_s^{\rm NP})}{1+y_s} \right] |S_{\ell\ell}|^2,
\end{equation}
where $\phi_s^{\rm NP}$ denotes a possible NP contribution to the $B^0_s$--$\bar B^0_s$ mixing
phase 
\begin{equation}
\phi_s = -2\beta_s + \phi_s^{\rm NP}. 
\end{equation}
Current experimental information from $B^0_s\to J/\psi \phi$ and decays with similar dynamics 
gives the following results \cite{peng-anat,Amh16,CKMFitter:2016}:
\begin{equation}\label{phi_s}
\phi_s = -0.030\pm 0.033 = -(1.72\pm 1.89)^{\circ}
\end{equation}
\begin{equation}\label{phis-NP}
\phi_s^{\rm NP}=0.007 \pm 0.033 =(0.4\pm 1.9)^{\circ},
\end{equation}
where we have used the SM value $\phi_s^{\rm SM}=-2\beta_s=-(2.12\pm0.04)^\circ$. Similar quantities 
can also be introduced for the $B^0_d\to\ell^+\ell^-$ decays, in analogy to the expressions given above.

\subsection{Scenario for the New Physics Analysis}\label{ssec:BSM}
A first analysis of the interplay between $\overline{R}^s_{\mu\mu}$ and 
${\cal A}^{\mu\mu}_{\Delta\Gamma_s}$ within specific models of physics beyond the SM was 
performed in Ref.~\cite{BFGK}, giving also a classification of various scenarios. In view of the new 
LHCb results for the $B^0_s\to\mu^+\mu^-$ mode, a very recent study was performed in 
Ref.~\cite{ANS}, highlighting also the importance of measuring ${\cal A}^{\mu\mu}_{\Delta\Gamma_s}$ 
for the search and exploration of NP effects. 

In order to illustrate NP effects, we shall consider a general scenario with no new sources of 
CP violation, i.e.\ real Wilson coefficients. This assumption could be explored with the help of the
CP-violating observables discussed in Refs.~\cite{Bsmumu-ADG,BFGK}. Moreover, we assume 
that we have flavour-universal Wilson coefficients, allowing us to introduce the notation
\begin{equation}
C_{10}\equiv C_{10}^{q,\ell\ell}, \quad  C_{10}'\equiv C_{10}^{q,\ell\ell'}
\end{equation}
\begin{equation}
C_{P}\equiv C_{P}^{q,\ell\ell}, \quad  C_{P}'\equiv C_{P}^{q,\ell\ell'}, \quad
C_{S}\equiv C_{S}^{q,\ell\ell}, \quad  C_{S}'\equiv C_{S}^{q,\ell\ell'},
\end{equation}
as well as
\begin{equation}
{\cal C}_{10}\equiv \frac{C_{10}-C_{10}'}{C_{10}^{\rm SM}}.
\end{equation}
Using data for rare $B\to K^{(*)}\mu^+\mu^-$ decays, the latter coefficient can be determined from 
experimental data (for a state-of-the-art analysis, see Ref.~\cite{ANSS}). Data for $B\to K^{(*)} e^+e^-$ 
modes allows us also to take a possible violation of Lepton Flavour Universality into account. 
As a working assumption, we shall use ${\cal C}_{10}=1$, which is consistent with the current rare 
$B$-decay data within the uncertainties, and corresponds to a picture of NP entering only through new
(pseudo)-scalar contributions, which is the key domain for the $B^0_q \to \ell^+\ell^-$ decays. We 
obtain then the following expressions:
\begin{equation}\label{P-expr-1}
P^q_{\ell\ell} = {\cal C}_{10}+\frac{M_{B_q}^2}{2 m_\ell}
\left(\frac{m_b}{m_b+m_q}\right)\left[\frac{C_P-C_P'}{C_{10}^{\rm SM}}\right]
\end{equation}
\begin{equation}\label{S-expr-1}
S^q_{\ell\ell}\equiv \sqrt{1-4\frac{m_\ell^2}{M_{B_q}^2}}
\frac{M_{B_q} ^2}{2 m_\ell}\left(\frac{m_b}{m_b+m_q}\right)
\left[\frac{C_S-C_S'}{C_{10}^{\rm SM}}\right],
\end{equation}
which will serve as the basis for our following discussion of NP effects. In particular, we shall not
assume any relation between the $P^q_{\mu\mu}$ and $S^q_{\mu\mu}$ coefficients, which 
typically arise in more general NP frameworks as well as in specific models \cite{BFGK,AGMC}.

In Ref.~\cite{ANS}, a scenario with heavy new degrees of freedom, which are linearly realized 
in the electroweak symmetry in the Higgs sector, and the feature of MFV was considered \cite{AGMC}, 
including the Minimal Supersymmetric Standard Model (MSSM) with MFV violation. In the latter case, 
the coefficients $C_{P}'$, $C_{S}'$ are suppressed by the mass ratio $m_q/m_b$, and the relation 
$C_S=-C_P$ holds. Moreover, these coefficients are proportional to the lepton mass $m_\ell$ 
(see also Refs.~\cite{BFGK,BHP}):
\begin{equation}\label{MSSM-rel}
C_S=m_\ell\tilde C_S, \quad C_P=m_\ell\tilde C_P,
\end{equation}
yielding
\begin{equation}\label{P-expr-MSSM}
P^q_{\ell\ell}|_{\rm MSSM}^{\rm MFV} = 1-\frac{M_{B_q}^2}{2}
\left(\frac{m_b}{m_b+m_q}\right)\left[\frac{\tilde C_S}{C_{10}^{\rm SM}}\right]\equiv1-A_q
\end{equation}
\begin{equation}\label{S-expr-MSSM}
S^q_{\ell\ell}|_{\rm MSSM}^{\rm MFV}= \sqrt{1-4\frac{m_\ell^2}{M_{B_q}^2}}
\frac{M_{B_q} ^2}{2}\left(\frac{m_b}{m_b+m_q}\right)
\left[\frac{\tilde C_S}{C_{10}^{\rm SM}}\right]=
\sqrt{1-4\frac{m_\ell^2}{M_{B_q}^2}}A_q,
\end{equation}
where $A_q$ does not depend on the lepton flavour $\ell$. If we neglect the $m_\ell^2/M_{B_q}^2$ 
term under the square root in (\ref{S-expr-MSSM}), both $P^q_{\ell\ell}$ and $S^q_{\ell\ell}$ 
are independent of the lepton mass in this scenario, implying that the ratios of branching ratios of the 
various $B^0_{s,d}\to\ell^+\ell^-$ decays are given as in the SM, up to ${\cal O}(m_\ell^2/M_{B_q}^2)$ 
corrections. In the case of the $B^0_{s,d}\to\tau^+\tau^-$ decays, these effects may have a 
sizeable impact, as we will discuss in Subsection~\ref{sec:NP-Btautau}.

The flavour-universal scenario introduced above offers an interesting general framework to explore 
NP effects in the $B^0_q\to \tau^+\tau^-$ and $B^0_q\to e^+e^-$ decays and to illustrate their potential
impact. But before focusing on these modes, let us first discuss the picture for the $B^0_q\to\mu^+\mu^-$ 
channels following from the current data.

\boldmath
\section{The Decays $B^0_s\to\mu^+\mu^-$ and $B^0_d\to\mu^+\mu^-$}\label{sec:Bsmumu}
\unboldmath
\subsection{Experimental Status}\label{subsec:BsmumuSM}
Using the results of Ref.~\cite{Bsmumu-SM} and rescaling them to the updated parameters 
collected in Table~\ref{tab:inputs}, we obtain the following SM branching ratios:
\begin{equation} \label{eq:BsmumuSM}
\overline{\mathcal{B}}(B_s \to \mu^+\mu^-)_\text{SM} = (3.57 \pm 0.16) \times 10^{-9}
\end{equation}
\begin{equation} \label{eq:BdmumuSM}
\overline{\mathcal{B}}(B_d \to \mu^+\mu^-)_\text{SM} = (1.02  \pm 0.06) \times 10^{-10}.
\end{equation}
On the experimental side, the LHCb collaboration has recently presented updated 
measurements of the $B^0_s\to\mu^+\mu^-$ and $B^0_d\to\mu^+\mu^-$ branching ratios \cite{LHCb-2017}:
\begin{equation}\label{eq:LHCb17Bs}
\overline{\mathcal{B}}(B_s \to \mu^+\mu^-)_\text{LHCb'17}  = 
\left(3.0\pm0.6^{+0.3}_{-0.2}\right) \times 10^{-9} 
\end{equation}
\begin{equation}\label{LHCb-Bdmumu}
\overline{\mathcal{B}}(B_d \to \mu^+\mu^-)_\text{LHCb'17} =
\left(1.5^{+1.2+0.2}_{-1.0-0.1}\right) \times 10^{-10}.
\end{equation}
The CP-averaged signal for $B^0_s \to \mu^+\mu^-$ has a statistical significance of $7.8\,\sigma$, while 
$B^0_d \to \mu^+\mu^-$ has a significance of $1.6\,\sigma$, corresponding to 
$\overline{\mathcal{B}}(B_d \to \mu^+\mu^-)<3.4\times 10^{-10}$ (95\% C.L.). These experimental 
results are consistent with the SM predictions within the uncertainties.  In 2013, the CMS 
collaboration reported the following result \cite{CMSmumu}: 
\begin{equation}\label{CMS-res}
\overline{\mathcal{B}}(B_s \to \mu^+\mu^-)_\text{CMS'13}=\left(3.0^{+1.0}_{-0.9}\right)\times 10^{-9},
\end{equation}
which corresponds to a signal with $4.3\, \sigma$ significance. The ATLAS collaboration presented 
the constraint $\overline{\mathcal{B}}(B_s \to \mu^+\mu^-)_\text{ATLAS'16}=(0.9^{+1.1}_{-0.8})
\times 10^{-9}$ in 2016 \cite{ATLAS:2016}, which we give for comparison. 
The combination of the results in Eqs.~(\ref{eq:LHCb17Bs}) and (\ref{CMS-res}) gives
\begin{equation} \label{eq:BsmumuExpComb}
\overline{\mathcal{B}}(B_s \to \mu^+\mu^-)_\text{LHCb'17+CMS} = (3.0 \pm 0.5 ) \times 10^{-9},
\end{equation}
where we have calculated the average by applying the procedure of the Particle Data Group (PDG)
\cite{PDG:2016}.

\begin{table}
\centering
\begin{tabular}{|c|c|c|c|}
\hline
Parameter& Value & Unit & Reference\\
\hline
\hline
$m_e$ &  $0.510 998 9461(31)\times 10^{-3}$   & GeV &\cite{Mohr:2014} \\
\hline
$m_{\mu}$ & $105.658 3745(24) \times 10^{-3}$     & GeV &\cite{Mohr:2014} \\
\hline
$m_{\tau}$ &  $1.77686 (12)$   & GeV &\cite{PDG:2016} \\
\hline
\hline
$m_d$     &   $(4.7^{+0.5}_{-0.4})\times 10 ^{-3}$   &  GeV   & \cite{PDG:2016} \\
\hline
$m_s$     &   $0.096^{+0.008}_{-0.004}$             &  GeV   &  \cite{PDG:2016} \\
\hline
$m_b$     & $4.18^{+0.04}_{-0.03}$                   &  GeV   &  \cite{PDG:2016} \\
\hline
$f_{B_s}$ &   $(228.4 \pm 3.7)\times 10^{-3}$       & GeV  &  \cite{Aoki:2016} \\
\hline
$f_{B_d}$ &   $(192.0 \pm 4.3)\times 10^{-3}$        & GeV  &  \cite{Aoki:2016}  \\
\hline
$f_{B_s}/f_{B_d}$ &   $1.201 \pm 0.016$        &   &  \cite{Aoki:2016}  \\
\hline
$\hat{B}_{B_d}$ &   $1.26 \pm 0.09$       &   &  \cite{Aoki:2016} \\
\hline
$\hat{B}_{B_s}$ &   $1.32 \pm 0.06$        &   &  \cite{Aoki:2016}  \\
\hline
$\tau_{B_s}$  &  $1.505\pm 0.005$ & ps  & \cite{Amh16}  \\
\hline
$\tau_{B_d}$ & $1.520 \pm 0.004$ & ps &  \cite{Amh16}  \\
\hline
$M_{B_s}$   &  $5.36682(22)$  &  GeV    & \cite{PDG:2016} \\
\hline
$M_{B_d}$   &  $5.27962(15)$  &  GeV   & \cite{PDG:2016} \\
\hline
\hline
$y_s$ & $0.0645\pm 0.0045$    &   &  \cite{Amh16}  \\
\hline
$\Delta M_s$      &$17.757 \pm 0.021$& $\hbox{ps}^{-1}$ &  \cite{Amh16} \\
\hline
$\Delta M_d$ & $0.5064 \pm 0.0019$ & $\hbox{ps}^{-1}$ &  \cite{Amh16}\\
\hline
$\phi_s$ & $-0.030\pm 0.033$ & &  \cite{Amh16}\\
\hline
\hline
$|V_{td}|$ & $0.008575^{+0.000076}_{-0.000098}$  & & \cite{CKMFitter:2016}\\
\hline
$|V_{ts}|$ & $0.04108^{+0.00030}_{-0.00057}$& &\cite{CKMFitter:2016}\\
\hline
$|V_{tb}|$ & $0.999119^{+0.000024}_{-0.000012}$& &\cite{CKMFitter:2016}\\
\hline
$\lambda$  & $0.22509^{+0.00029}_{-0.00028}$     & &\cite{CKMFitter:2016}\\
\hline
$\beta_s$ & $0.01852^{+0.00032 }_{-0.00032}$ & & \cite{CKMFitter:2016}\\
\hline
\end{tabular}
\caption{Input parameters used in the numerical evaluations of this paper.\label{tab:inputs}}
\end{table}

The LHCb collaboration has very recently reported a first measurement of the effective lifetime
of the $B^0_s\to\mu^+\mu^-$ decay \cite{LHCb-2017}:
\begin{equation} \label{eq:tauEffmumu}
\tau^s_{\mu\mu} = \left[2.04 \pm 0.44 ({\rm stat}) \pm 0.05 ({\rm syst}) \right]\hbox{ps}.
\end{equation}
Using the expression
\begin{equation}
\frac{\tau^s_{\mu\mu}}{\tau_{B_s}}=\frac{1+2y_s\mathcal{A}_{\Delta\Gamma_s}^{\mu\mu} +y_s^2}{(1+y_s
\mathcal{A}_{\Delta\Gamma_s}^{\ell\ell} )(1-y_s^2)}
\end{equation}
with Eq.~(\ref{eq:ADGSM}) and the numerical inputs in Table~\ref{tab:inputs}, we obtain the SM prediction
\begin{equation}
\tau^s_{\mu\mu}|_{\rm SM}=\frac{\tau_{B_s}}{1-y_s} = (1.61 \pm 0.01) \, \hbox{ps}.
\end{equation}
It agrees with the LHCb value, although the experimental uncertainties are too large to draw further
conclusions. Using Eq.\ (\ref{eq:aDGLifetime}), we may convert Eq.~(\ref{eq:tauEffmumu}) into 
\begin{equation}\label{ADG-exp}
\mathcal{A}_{\Delta\Gamma_s}^{\mu\mu} = 8.24 \pm 10.72,
\end{equation}
where the error is fully dominated by the huge uncertainty on the effective lifetime 
$\tau^s_{\mu\mu}$. As we have the model-independent relation 
\begin{equation}\label{ADG-range}
-1\leq\mathcal{A}_{\Delta\Gamma_s}^{\mu\mu}\leq+1, 
\end{equation}
it will be crucial to improve the experimental precision for this observable in the future data 
taking at the LHC.

\boldmath
\subsection{General Constraints on New Physics}\label{sec:NewPhysics}
\unboldmath
\begin{figure}
\centering
\includegraphics[height=8cm]{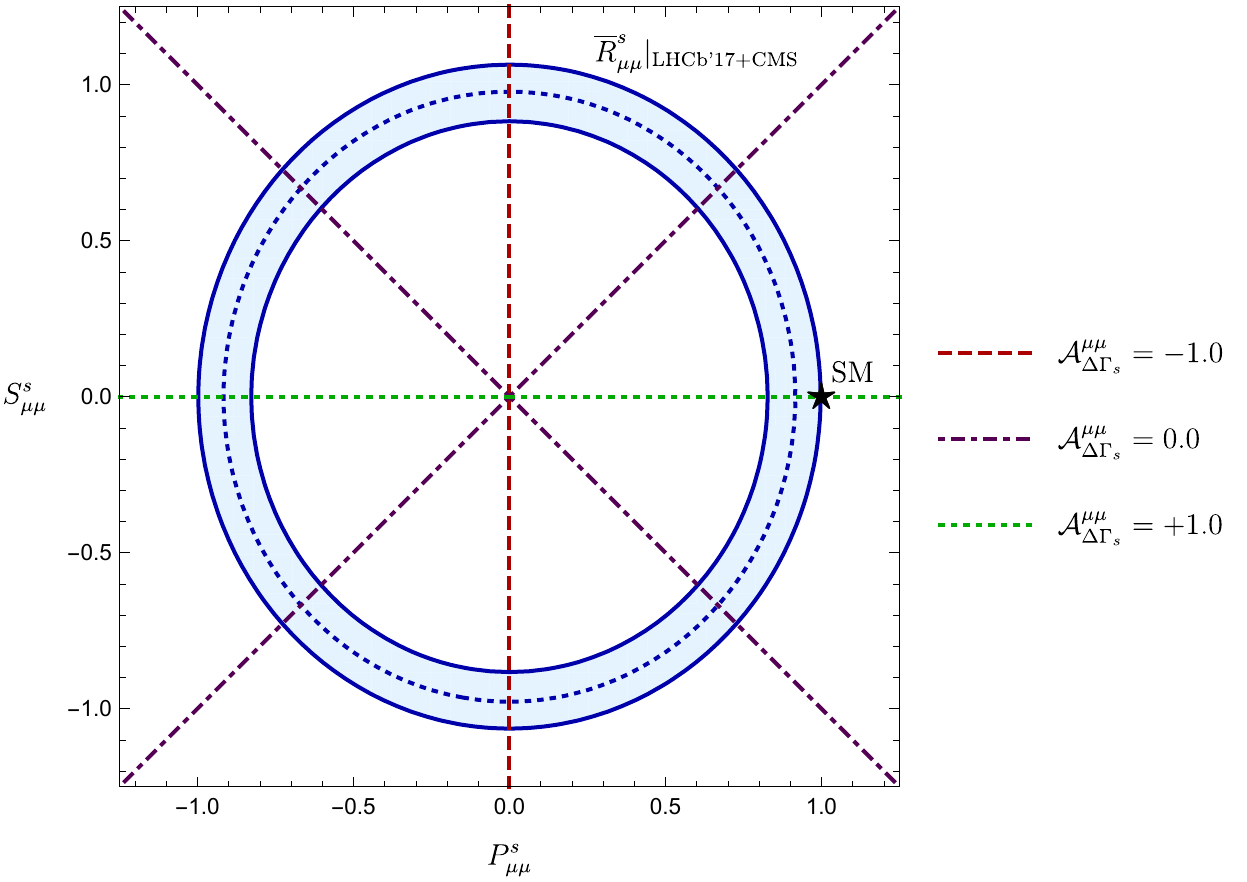}
\caption{Constraints in the $P^s_{\mu\mu}$--$S^s_{\mu\mu}$ plane, assuming real coefficients 
with trivial CP-violating NP phases $\varphi_{P_s}^{\mu\mu}, \varphi_{S_s}^{\mu\mu}\in\{0,\pi\}$.
The blue circular band corresponds to the current experimental information for 
$\overline{R}^s_{\mu\mu}$. A future precise measurement of the effective lifetime $\tau^s_{\mu\mu}$ 
and thus $\mathcal{A}_{\Delta\Gamma_s}^{\mu\mu}$, as illustrated by the dotted and dashed lines, will pin 
down values for $P^s_{\mu\mu}$ and $S^s_{\mu\mu}$ up to discrete ambiguities.}\label{fig:SP}
\end{figure}

Let us first have a look at the $B^0_s\to\mu^+\mu^-$ decay observables. Using 
Eqs.~(\ref{eq:BsmumuSM}) and (\ref{eq:BsmumuExpComb}), we can determine the ratio
$\overline{R}^s_{\mu\mu}$ from Eq.~(\ref{Rellells}):
\begin{eqnarray}\label{eq:Rbarmumuval}
\left.\overline{R}^s_{\mu\mu}\right|_\text{LHCb'17+CMS} &=& 0.84 \pm 0.16.
\end{eqnarray}
Assuming that we have no new CP-violating phases in $P^s_{\mu\mu}$ and $S^s_{\mu\mu}$, as in the
NP model introduced in Subsection~\ref{ssec:BSM}, expression (\ref{eq:RBarll}) reduces to
\begin{equation}\label{eq:Rbarmumutheo}
\overline{R}^s_{\mu\mu} = \left[\frac{1+y_s\cos\phi_s^{\rm NP}}{1+y_s}\right]|P^s_{\mu\mu}|^2 + 
\left[\frac{1-y_s\cos\phi_s^{\rm NP}}{1+y_s}\right]|S^s_{\mu\mu}|^2.
\end{equation}
Using the experimental value of $\phi_s^{\rm NP}$ in Eq.~(\ref{phi_s}) we get
\begin{equation}\label{cosNP}
\cos\phi_s^{\rm NP} = 1.0000(2),
\end{equation}
which allows us to convert Eq.~(\ref{eq:Rbarmumuval}) into a circular band in the 
$|P^s_{\mu\mu}|$--$|S^s_{\mu\mu}|$ plane. 

The observable $\mathcal{A}_{\Delta\Gamma_s}^{\mu\mu}$ provides another constraint in 
this parameter space. Assuming real coefficients  $P^s_{\mu\mu}$ and $S^s_{\mu\mu}$, 
Eq.~(\ref{ADG}) yields
\begin{equation}\label{eq:ADGellell}
\mathcal{A}_{\Delta\Gamma_s}^{\mu\mu} = 
\cos\phi_s^{\rm NP}\left[\frac{(P^s_{\mu\mu})^2 - (S^s_{\mu\mu})^2}{(P^s_{\mu\mu})^2 + 
(S^s_{\mu\mu})^2}\right],
\end{equation}
fixing a straight line in the $P^s_{\mu\mu}$--$S^s_{\mu\mu}$ plane through the measured value of 
$\mathcal{A}_{\Delta\Gamma_s}^{\mu\mu}$. Interestingly, as the NP phases phases enter as 
$2 \varphi_{P_s}^{\mu\mu}$ and $2 \varphi_{S_s}^{\mu\mu}$ in Eqs.~(\ref{ADG}) and (\ref{eq:RBarll}), 
we cannot reveal minus signs of $P^s_{\mu\mu}$ and $S^s_{\mu\mu}$, which correspond to 
$\varphi_{P_s}^{\mu\mu}, \varphi_{S_s}^{\mu\mu}=\pi$, leading to terms of $2\pi$ in the arguments
of the relevant trigonometric functions, thereby leaving them unchanged. Consequently, 
$\overline{R}^s_{\mu\mu}$ and $\mathcal{A}_{\Delta\Gamma_s}^{\mu\mu}$ allow us to determine 
$P^s_{\mu\mu}$ and $S^s_{\mu\mu}$ up to discrete ambiguities:
\begin{displaymath}
|P^s_{\mu\mu}|=\sqrt{\frac{(1+y_s)(\cos\phi_s^{\rm NP}+\mathcal{A}_{\Delta\Gamma_s}^{\mu\mu})
\overline{R}^s_{\mu\mu}}{(1+y_s\cos\phi_s^{\rm NP})(\cos\phi_s^{\rm NP}+
\mathcal{A}_{\Delta\Gamma_s}^{\mu\mu})+(1-y_s\cos\phi_s^{\rm NP})(\cos\phi_s^{\rm NP}-
\mathcal{A}_{\Delta\Gamma_s}^{\mu\mu})}}
\end{displaymath}
\begin{equation}\label{P-det}
=\sqrt{\frac{1}{2}\left(1+y_s\right)\left[\frac{1+\mathcal{A}_{\Delta\Gamma_s}^{\mu\mu}}{1+y_s
\mathcal{A}_{\Delta\Gamma_s}^{\mu\mu}}\right]\overline{R}^s_{\mu\mu}}
\end{equation}
\vspace*{0.5truecm}
\begin{displaymath}
|S^s_{\mu\mu}|=\sqrt{\frac{(1+y_s)(\cos\phi_s^{\rm NP}-\mathcal{A}_{\Delta\Gamma_s}^{\mu\mu})
\overline{R}^s_{\mu\mu}}{(1+y_s\cos\phi_s^{\rm NP})(\cos\phi_s^{\rm NP}+
\mathcal{A}_{\Delta\Gamma_s}^{\mu\mu})+(1-y_s\cos\phi_s^{\rm NP})(\cos\phi_s^{\rm NP}-
\mathcal{A}_{\Delta\Gamma_s}^{\mu\mu})}}
\end{displaymath}
\begin{equation}\label{S-det}
=\sqrt{\frac{1}{2}\left(1+y_s\right)\left[\frac{1-\mathcal{A}_{\Delta\Gamma_s}^{\mu\mu}}{1+y_s
\mathcal{A}_{\Delta\Gamma_s}^{\mu\mu}}\right]\overline{R}^s_{\mu\mu}},
\end{equation}
where we have also given the simplified expressions for $\cos\phi_s^{\rm NP}=1$. In Fig.~\ref{fig:SP}, 
we illustrate the resulting situation in the $P^s_{\mu\mu}$--$S^s_{\mu\mu}$ plane, showing both
the circular band arising from the current experimental value of $\overline{R}^s_{\mu\mu}$ and the impact
of a future measurement of the $\mathcal{A}_{\Delta\Gamma_s}^{\mu\mu}$ observable. 

In order to test the SM with the $B^0_d\to\mu^+\mu^-$ decay, it is advantageous to 
consider the ratio of its branching ratio and the one of $B^0_s\to\mu^+\mu^-$ \cite{BF-rev}. 
We obtain the following general expression:
\begin{eqnarray}\label{BRds-rat}
\frac{\overline{\mathcal{B}}(B_d \to \mu^+\mu^-)}{\overline{\mathcal{B}}(B_s \to \mu^+\mu^-)}&=&
\frac{\tau_{B_d}}{\tau_{B_s}}\left[\frac{1-y_s^2}{1-y_d^2}\right]
\left[\frac{1+{\cal A}^{\mu\mu}_{\Delta\Gamma_d}y_d}{1+{\cal A}^{\mu\mu}_{\Delta\Gamma_s}y_s}\right]
\left[\frac{|P^d_{\mu\mu}|^2 + |S^d_{\mu\mu}|^2}{|P^s_{\mu\mu}|^2 + |S^s_{\mu\mu}|^2}\right] 
\times \nonumber\\
&\times& \frac{M_{B_d}}{M_{B_s}}
\sqrt{\frac{1-4\left(m_\mu^2/M_{B_d}^2\right)}{1-4\left(m_\mu^2/M_{B_s}^2\right)}}
\left(\frac{f_{B_d}}{f_{B_s}}\right)^2\left|\frac{V_{td}}{V_{ts}}\right|^2.
\end{eqnarray}

The CKM factor $|V_{td}/V_{ts}|$ is required to utilize this ratio and has to be determined in a
way that is robust with respect to the impact of NP effects. Assuming the unitarity of the
CKM matrix, it can be extracted from the the length 
\begin{equation}
R_t\equiv\sqrt{(1-\bar\rho)^2+\bar\eta^2}=\frac{1}{\lambda}\left|\frac{V_{td}}{V_{cb}}\right|
\end{equation}
of the Unitarity Triangle (UT) as $|V_{cb}|=|V_{ts}|+{\cal O}(\lambda^2)$. Here $\lambda\equiv|V_{us}|$ 
is the Wolfenstein parameter \cite{Wol83}, and $(\bar\rho,\bar\eta)$ describes the apex of the UT 
in the complex plane \cite{BLO}. Taking subleading corrections in $\lambda$ into account and 
employing the UT side
\begin{equation}
R_b\equiv\sqrt{\bar\rho^2+\bar\eta^2}
=\left(1-\frac{\lambda^2}{2}\right)\frac{1}{\lambda}\left|\frac{V_{ub}}{V_{cb}}\right|,
\end{equation}
we obtain
\begin{equation}\label{CKM-rat}
\left|\frac{V_{td}}{V_{ts}}\right|=\lambda\left[\frac{\sqrt{(1-R_b\cos\gamma)^2+
(R_b\sin\gamma)^2}}{1-1/2(1-2R_b\cos\gamma)\lambda^2}\right]+{\cal O}(\lambda^5),
\end{equation}
where $\gamma$ is the angle between $R_b$ and the real axis. 

Using pure tree decays of the kind $B\rightarrow D^{(*)}K^{(*)}$ \cite{gw,ADS}, $\gamma$ can 
be determined in a theoretically clean way (for an overview, see \cite{FR-gam}). The current
experimental value is given as follows \cite{CKMFitter:2016}:
\begin{equation}\label{eq:CKMGamma}
\gamma=(72.1_{-5.8}^{+5.4})^\circ.
\end{equation}
In the future, thanks to Belle II \cite{Belle-II} and the LHCb upgrade \cite{LHCbup}, the uncertainty 
for $\gamma$ is expected to be reduced to the $1^\circ$ level.

\begin{figure}
\centering
\includegraphics[height=5cm]{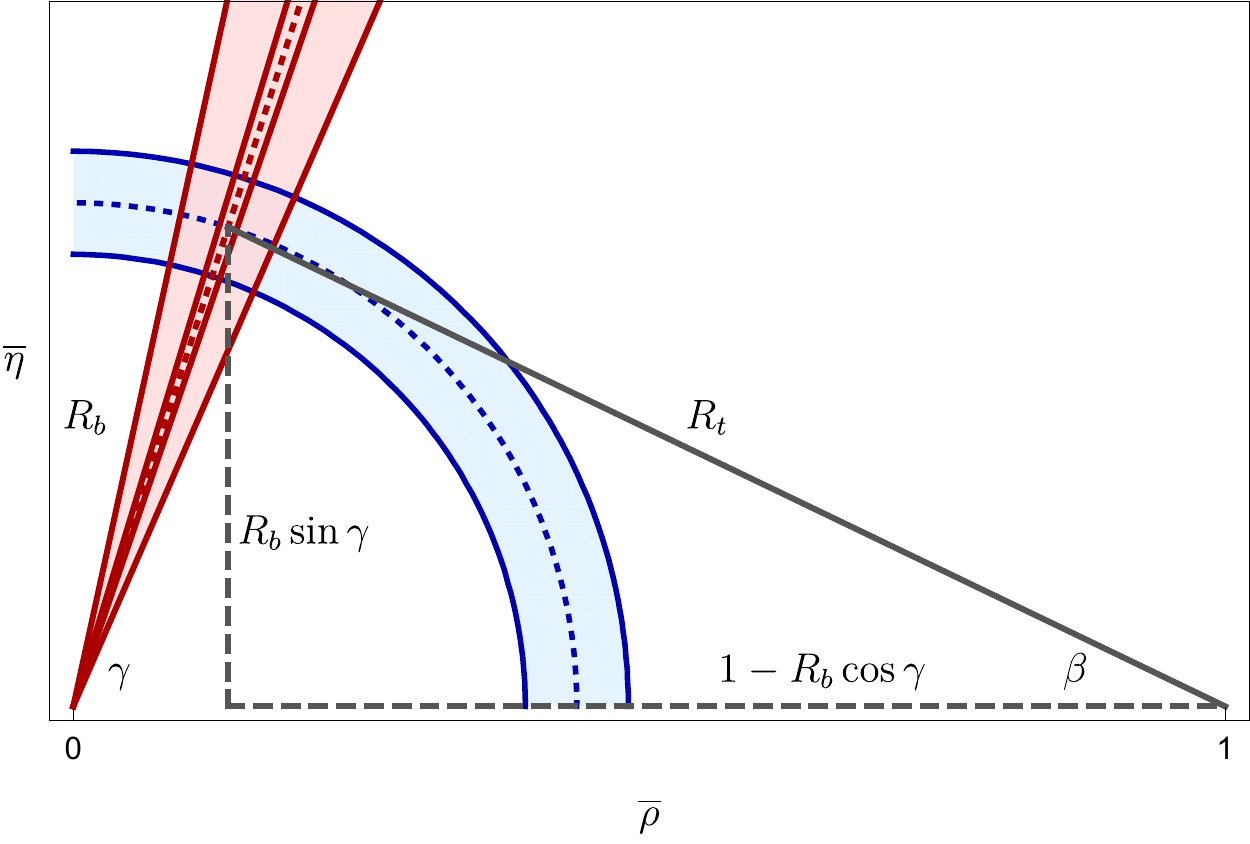}
\includegraphics[height=5cm]{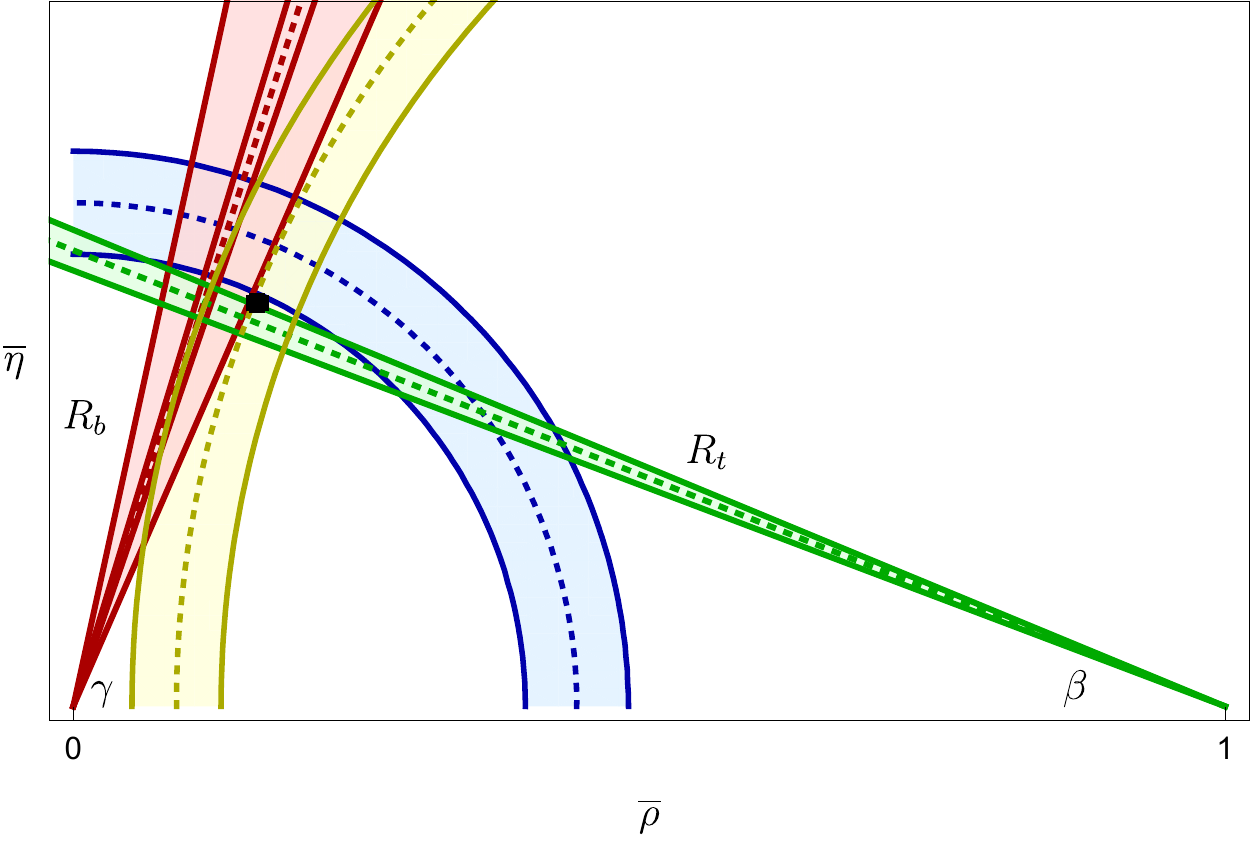}
\caption{Illustration of the Unitarity Triangle in the $\bar\rho$--$\bar\eta$ plane. The left 
panel illustrates the determination of the side $R_t$, where the blue circular band for $R_b$ corresponds 
to the average in Eq.\ (\ref{Rb-av}) and the wide red sector to the current value of $\gamma$ in 
Eq.\ (\ref{eq:CKMGamma}); the narrow red sector illustrates the future improvement to $1^{\circ}$ 
precision. In the right panel, we add constraints from $\Delta M_s/\Delta M_d$ and CP violation in 
$B^0_d\to J/\psi K_{\rm S}$, and show the small black region for the apex following from the 
comprehensive fit of Ref.~\cite{CKMFitter:2016}.}\label{fig:UT}
\end{figure}

Concerning the $R_b$ side, it can be determined with the help of $|V_{ub}|$ and $|V_{cb}|$ 
extracted from analyses of exclusive and inclusive semileptonic $B$ decays (for an overview, 
see the corresponding review in Ref.~\cite{PDG:2016}). The current status can be summarized as 
\begin{equation}\label{Rbincl-excl}
R_b|_{\rm incl} = 0.46\pm0.03, \quad R_b|_{\rm excl} =0.41\pm0.02,
\end{equation}
with the average
\begin{equation}\label{Rb-av}
R_b = 0.44\pm0.04.
\end{equation}

The determinations of $\gamma$ and $R_b$ using pure tree decays are very robust with respect 
to NP effects. Consequently, they allow us to determine the ratio in Eq.~(\ref{CKM-rat}) in a way 
that is also very robust concerning NP contributions, serving as the reference value for the analysis 
of Eq.~(\ref{CKM-rat}). The current data with the average value of $R_b$ in Eq.\ (\ref{Rb-av}) give
\begin{equation}
\left|\frac{V_{td}}{V_{ts}}\right|=0.220\pm0.010,
\end{equation}
where the error may be reduced to $0.002$ in the Belle II and LHCb upgrade era. We have illustrated 
the resulting situation for the UT in the complex plane in the left panel of Fig.~\ref{fig:UT}. Thanks to 
the specific shape of the UT, we observe that the uncertainty of $R_t$ is fully governed by $\gamma$, 
while the uncertainty of $R_b$ has a minor impact. Consequently, also the discrepancy between the 
inclusive and exclusive determinations in Eq.~(\ref{Rbincl-excl}) has fortunately a negligible effect in this case. 
It is impressive to see the impact of the future extraction of $\gamma$, allowing a very precise 
determination of $R_t$. For completeness, in the right panel of Fig.~\ref{fig:UT}, we show other 
constraints in the $\bar\rho$--$\bar\eta$ plane following from $\Delta M_s/\Delta M_d$ and the 
determination of the CKM angle $\beta=(21.6\pm 0.9)^{\circ}$ through CP violation in
$B_d^0 \to J/\psi K_{\rm S}^0$ decays, taking penguin effects into account \cite{peng-anat}. 
For comprehensive analyses of the UT, the reader is referred to Refs.~\cite{CKMFitter:2016,UTfit,BlBu}.

Using the CKM factor $|V_{td}/V_{ts}|$ as determined through Eq.\ (\ref{CKM-rat}), 
we may convert the measured ratio 
of the $B^0_{s,d}\to\mu^+\mu^-$ branching ratios into the following parameter:
\begin{eqnarray}
U^{ds}_{\mu\mu}\equiv\sqrt{\frac{|P^d_{\mu\mu}|^2 + |S^d_{\mu\mu}|^2}{|P^s_{\mu\mu}|^2 + 
|S^s_{\mu\mu}|^2}}
&=&\Biggl[\frac{\tau_{B_s}}{\tau_{B_d}}\left[\frac{1-y_d^2}{1-y_s^2}\right]
\left[\frac{1+{\cal A}^{\mu\mu}_{\Delta\Gamma_s}y_s}{1+{\cal A}^{\mu\mu}_{\Delta\Gamma_d}y_d}\right]
\frac{M_{B_s}}{M_{B_d}}\sqrt{\frac{1-4\left(m_\mu^2/M_{B_s}^2\right)}{1-4\left(m_\mu^2/M_{B_d}^2\right)}} 
\times \nonumber\\
&\times& \left(\frac{f_{B_s}}{f_{B_d}}\right)^2
\left|\frac{V_{ts}}{V_{td}}\right|^2
\left[\frac{\overline{\mathcal{B}}(B^0_d \to \mu^+\mu^-)}{\overline{\mathcal{B}}(B^0_s \to \mu^+\mu^-)}
\right]\Biggr]^{1/2},\label{U-def}
\end{eqnarray} 
which satisfies
\begin{equation}\label{Uds-SM}
U^{ds}_{\mu\mu}|_{\rm SM}=1.
\end{equation}
For NP models with MFV, which are characterized by universal short-distance functions, 
we have -- with excellent accuracy -- also a value of $U^{ds}_{\mu\mu}$ around one. 
A tiny difference may arise from the following small differences \cite{PDG:2016}:
\begin{eqnarray}\label{eq:massdiff}
M_{B_s}-M_{B_d}&=&(0.0872\pm 0.0003)~\text{GeV}\nonumber\\
\frac{m_b}{m_b + m_d}-\frac{m_b}{m_b + m_s}&=&\frac{m_b(m_s-m_d)}{(m_b+m_d)(m_b+m_s)}
= 0.021\pm 0.002.
\end{eqnarray}
We shall return to this parameter within our general flavour-universal NP scenario in the next 
Subsection (see Eq.~(\ref{Udet})). 

The current data give
\begin{equation}
U^{ds}_{\mu\mu} = 1.26\pm 0.49,
\end{equation}
where the error is unfortunately too large to draw conclusions. At the end of the LHCb upgrade, 
corresponding to 50 fb$^{-1}$ of integrated luminosity, LHCb expects to determine the ratio
$\overline{\mathcal{B}}(B^0_d \to \mu^+\mu^-)/\overline{\mathcal{B}}(B^0_s \to \mu^+\mu^-)$ with 
a precision at the 35\% level  \cite{LHCbup}. Assuming a future measurement of $\tau_{\mu\mu}^s$,
which determines $\mathcal{A}_{\Delta\Gamma_s}^{\mu\mu}$ through Eq.\ (\ref{eq:aDGLifetime}),
with a precision of $5\%$ \cite{Bsmumu-ADG} and a reduction in the uncertainty of $\gamma$ 
to $1^{\circ}$ would yield
\begin{equation}
U^{ds}_{\mu\mu} = 1.26\pm 0.23,
\end{equation}
which would still not allow a stringent test in view of the significant uncertainty. 
We can straightforwardly generalize the observable $U^{ds}_{\mu\mu}$ defined in Eq.\ 
(\ref{U-def}) to neutral $B^0_{d,s}$ decays with $\tau^+\tau^-$ and $e^+e^-$ leptons 
in the final state, as discussed below.

Should future measurements find a result for $U^{ds}_{\mu\mu}$ consistent with 1, thereby 
supporting the picture of the SM and models with MFV, we could extract the $SU(3)$-breaking 
ratio of ``bag" parameters 
describing $B^0_q$--$\bar B^0_q$ mixing from the following relation (see also Ref.~\cite{buras-rel}):
\begin{equation}\label{Bag-det}
\frac{\hat B_{B_s}}{\hat B_{B_d}}=\frac{\tau_{B_s}}{\tau_{B_d}}\left[\frac{1-y_d^2}{1-y_s^2}\right]
\left[\frac{1+{\cal A}^{\mu\mu}_{\Delta\Gamma_s}y_s}{1+{\cal A}^{\mu\mu}_{\Delta\Gamma_d}y_d}\right]
\left[\frac{\overline{\mathcal{B}}(B^0_d \to \mu^+\mu^-)}{\overline{\mathcal{B}}(B^0_s \to \mu^+\mu^-)}
\right]\left[\frac{\Delta M_s}{\Delta M_d}\right],
\end{equation}
allowing -- in principle -- an interesting test of lattice QCD. An agreement between experiment and theory 
would also support the lattice QCD calculation of the decay constants $f_{B_q}$, which are key inputs 
for the SM branching ratios. However, even in the LHCb upgrade era, we would only get a precision for 
Eq.\ (\ref{Bag-det}) at the level of $\pm 37 \%$, while current lattice QCD calculations 
give the following picture \cite{Aoki:2016}:
\begin{equation}\label{eq:latticebag}
\frac{\hat B_{B_s}}{\hat B_{B_d}}\Bigl|_{\rm Lattice}= 1.05 \pm 0.09.
\end{equation}
In order to  determine  the ratio of the bag parameters  using Eq.~(\ref{Bag-det})
with the same relative error as the one achieved by the current lattice calculations in 
Eq.~(\ref{eq:latticebag}), the measurement of  
$\overline{\mathcal{B}}(B^0_{d}\to\mu^+\mu^-)/\overline{\mathcal{B}}(B^0_{s}\to\mu^+\mu^-)$ 
should reach the $6\%$ precision while having the $5\%$ error in the measurement of the effective 
lifetime as for the LHCb upgrade. In this future scenario, we would be able to achieve a precision 
of $\pm 0.06$ for the determination of the observable $U^{ds}_{\mu\mu}$, which would be interesting
territory to search for signals of physics from beyond the SM.

\subsection{New Physics Benchmark Scenario}\label{subsec:BsmumuNP}
\begin{figure}
\centering
\includegraphics[height=8cm]{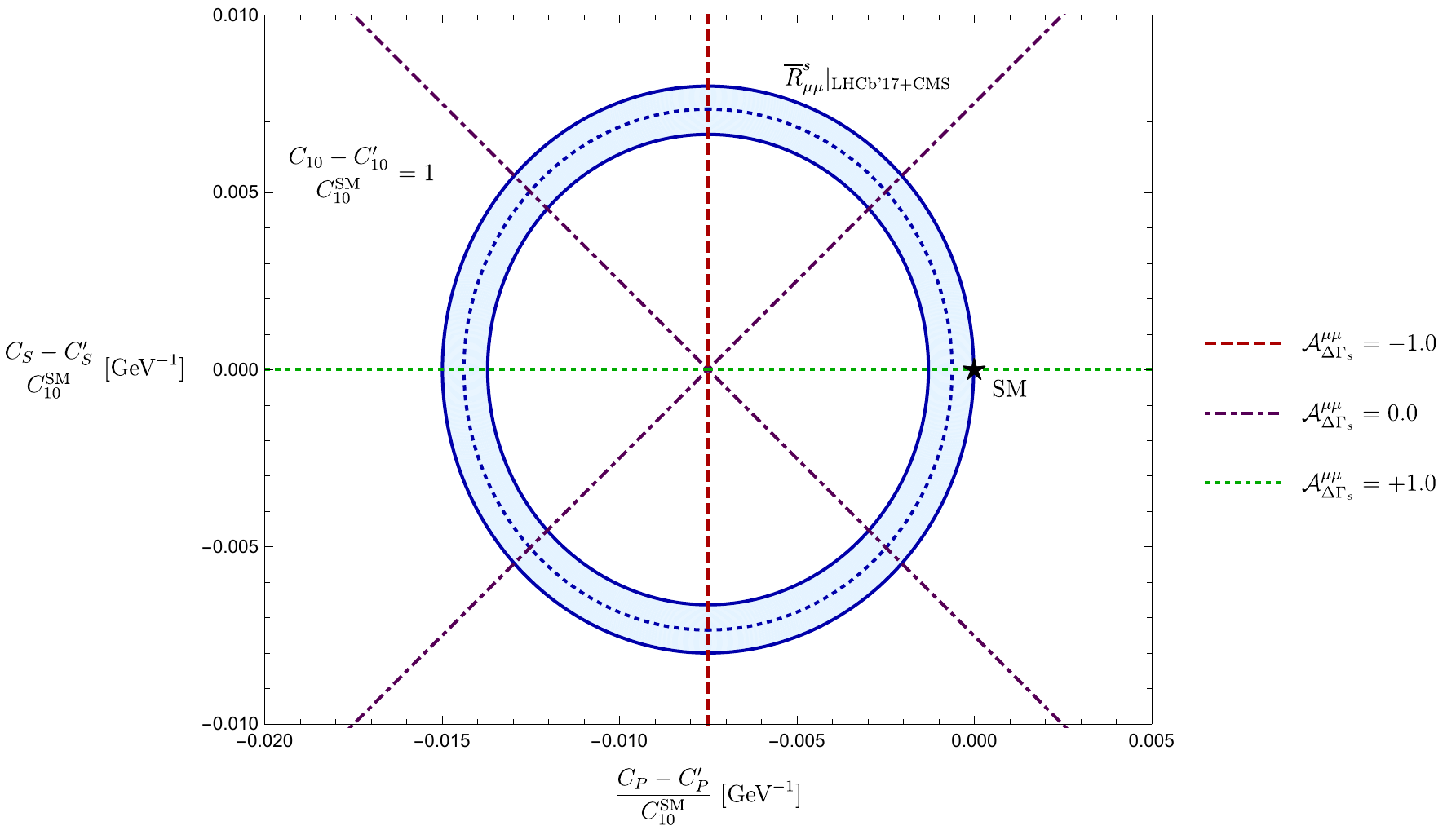}
\caption{Correlation between the ratios of Wilson coefficients in Eq.\ (\ref{WC-det}) for ${\cal C}_{10}=1$. 
We show the $1\sigma$ range corresponding to the measured value of $\overline{R}^s_{\mu\mu}$ in 
Eq.\ (\ref{eq:Rbarmumuval}), which is described by the blue circular band, and give lines for 
various values of $\mathcal{A}_{\Delta\Gamma_s}^{\mu\mu}$.}\label{fig:CP-CS}
\end{figure}

Let us now consider the situation in the general flavour-universal NP scenario introduced in 
Subsection~\ref{ssec:BSM}, which is characterized by Eqs.~(\ref{P-expr-1}) and (\ref{S-expr-1}),
and assume that not only the ratio $\overline{R}^s_{\mu\mu}$ but also the observable 
$\mathcal{A}_{\Delta\Gamma_s}^{\mu\mu}$ has been measured. Using Eqs.\ (\ref{P-det}) and 
(\ref{S-det}), we may then determine the coefficients $|P^s_{\mu\mu}|$ and $|S^s_{\mu\mu}|$, 
respectively, which allow us to extract the following ratios of short-distance coefficients:
\begin{equation}\label{CP-det}
\frac{C_P-C_P'}{C_{10}^{\rm SM}}=\frac{2 m_\mu}{M_{B_s}^2}\left(\frac{m_b+m_s}{m_b}\right)
\left[P^s_{\mu\mu} - {\cal C}_{10}\right]
\end{equation}
\begin{equation}\label{CS-det}
\frac{C_S-C_S'}{C_{10}^{\rm SM}} = \frac{2 m_\mu}{M_{B_s}^2}\left(\frac{m_b+m_s}{m_b}\right)
\frac{S^s_{\mu\mu}}{\sqrt{1-4\frac{m_\mu^2}{M_{B_q}^2}}}.
\end{equation} 
Since we can only determine the absolute values of $P^s_{\mu\mu}$ and $S^s_{\mu\mu}$, which
are real in our scenario, we have also to allow for negative values. We illustrate the corresponding
situation in Fig.~\ref{fig:CP-CS}. As the current measurement of $\mathcal{A}_{\Delta\Gamma_s}^{\mu\mu}$ 
in Eq.~(\ref{ADG-exp}) does not yet provide a useful constraint, we vary this observable within its
general range in Eq.\ (\ref{ADG-range}), yielding
\begin{equation}\label{WC-det}
-0.02 ~\text{GeV}^{-1}\leq \frac{C_P-C_P'}{C_{10}^{\rm SM}}\leq 0.00~\text{GeV}^{-1} , 
\quad -0.01~\text{GeV}^{-1} \leq \frac{C_S-C_S'}{C_{10}^{\rm SM}} \leq 0.01~\text{GeV}^{-1}.
\end{equation}
Once the observable $\mathcal{A}_{\Delta\Gamma_s}^{\mu\mu}$ has been measured with higher 
precision, these allowed ranges can be narrowed down correspondingly. In Ref.~\cite{AGMC}, 
constraints on similar coefficients were obtained.

\begin{figure}
\centering
\includegraphics[height=6.8cm]{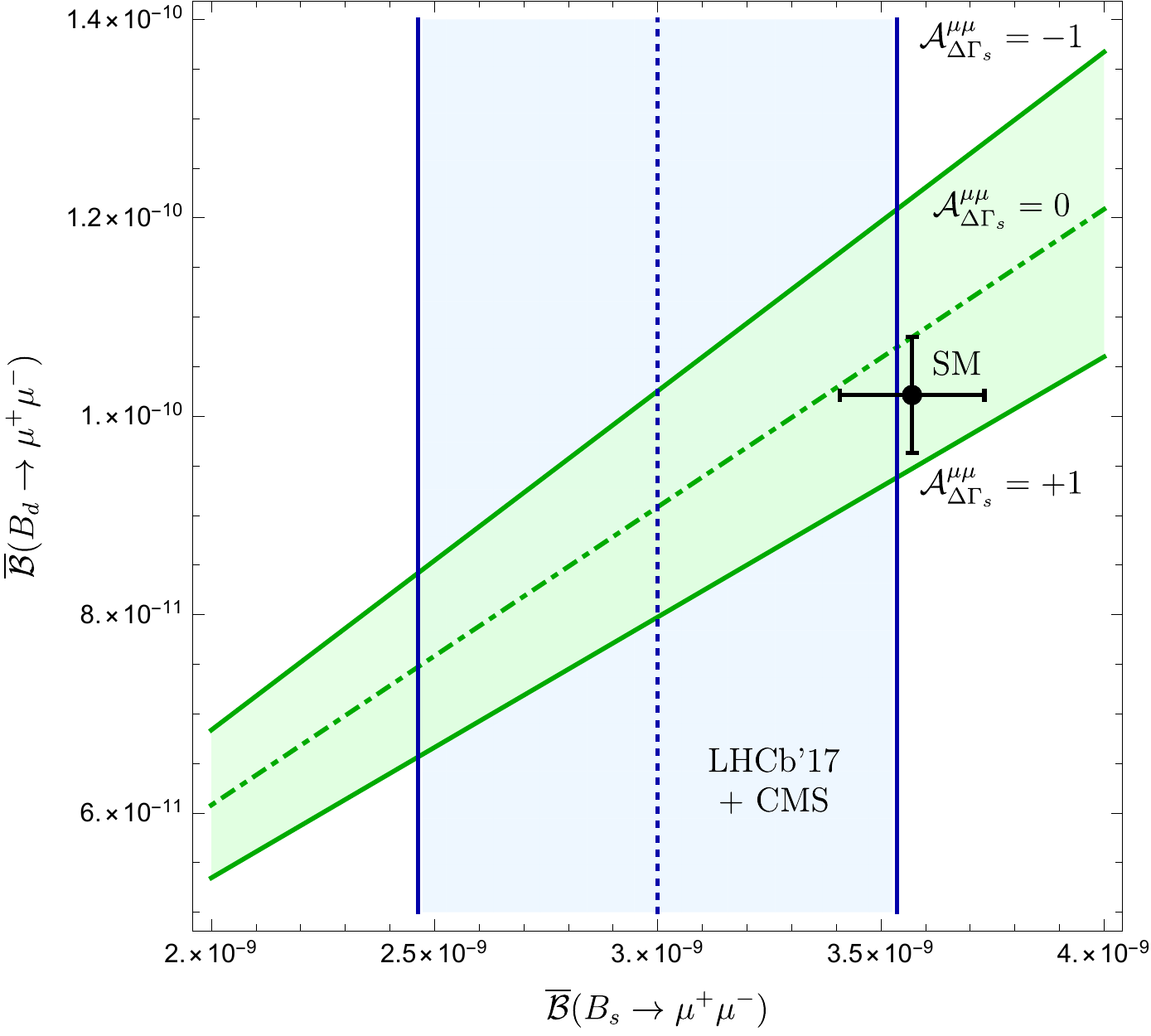}
\includegraphics[height=6.8cm]{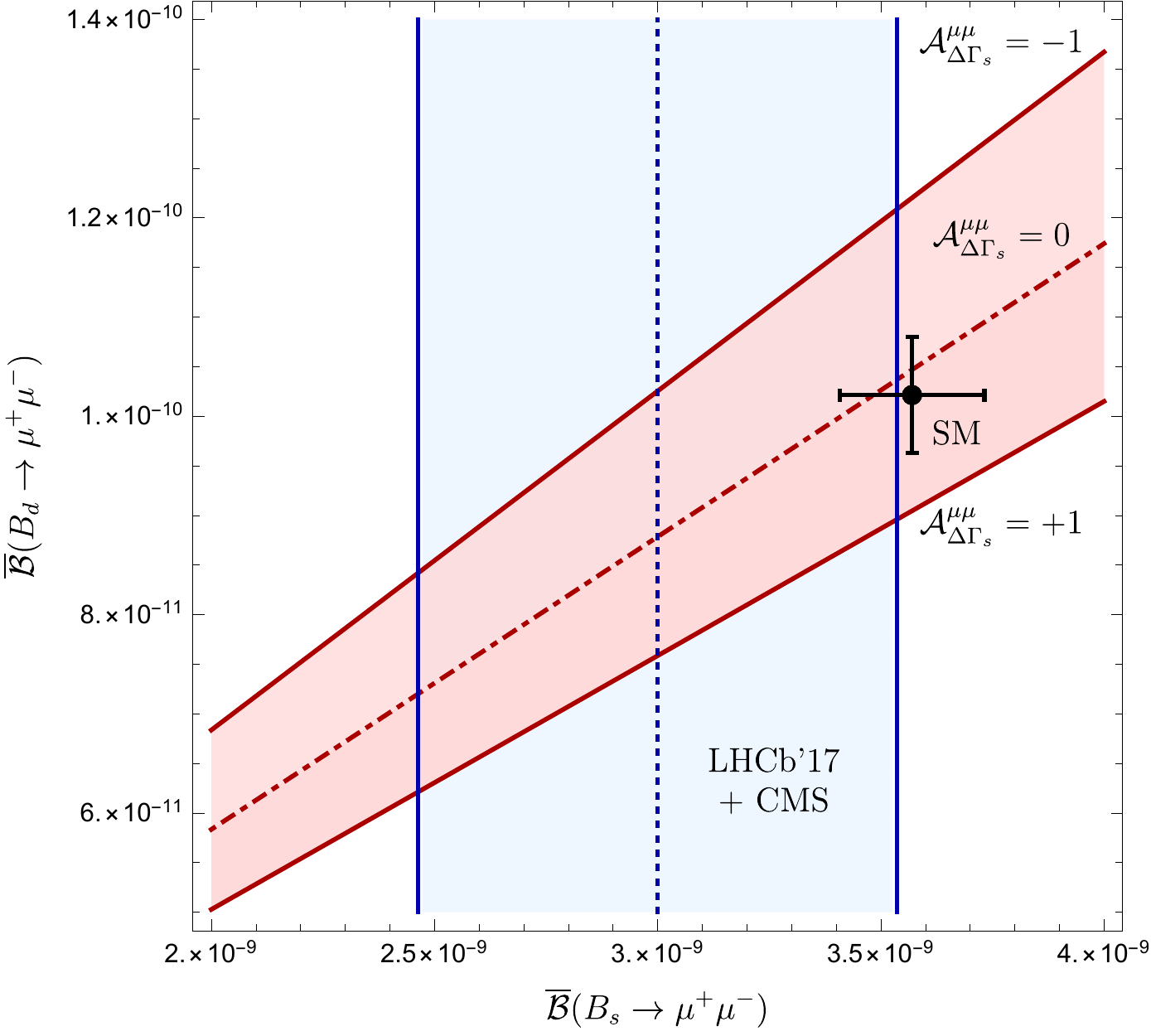}
\caption{Correlation between the branching ratios $\overline{\mathcal{B}}(B_d \rightarrow \mu^+\mu^-)$  
and $\overline{\mathcal{B}}(B_s \rightarrow \mu^+\mu^-)$ in the considered flavour-universal
NP scenario. The vertical blue band gives the $1\,\sigma$ range for the $B_s^0 \to \mu^+\mu^-$ branching ratio, while the green and red regions are obtained by varying $\mathcal{A}_{\Delta\Gamma_s}^{\mu\mu}$ between $-1$ and $+1$. The panels on the left- and right-hand sides 
correspond to positive and negative solutions for $P_{\mu\mu}^s$, 
respectively.}\label{fig:showbrdeebrBarsmumu}
\end{figure}

The coefficients in Eqs.~(\ref{CP-det}) and (\ref{CS-det}), with the corresponding ranges in 
Eq.\ (\ref{WC-det}), may now be used to study correlations with the other $B^0_{s,d}\to\ell^+\ell^-$ 
decays. Following these lines, we obtain the correlation between the branching ratios of the 
$B^0_d\to\mu^+\mu^-$ and $B^0_s\to\mu^+\mu^-$ decays shown in 
Fig.~\ref{fig:showbrdeebrBarsmumu}, corresponding to the numerical range
\begin{equation}\label{BR-Bd-sol}
0.66 \times 10^{-10}\leq\overline{\mathcal{B}}(B_d \rightarrow \mu^+\mu^-) \leq 1.14  \times 10^{-10},
\end{equation}
with
\begin{equation}
0.65 \leq \overline{R}_{\mu\mu}^d \leq 1.11,
\end{equation}
which is consistent with the LHCb result in Eq.~(\ref{LHCb-Bdmumu}). Finally, we obtain 
\begin{equation}\label{Udet}
0.97 \leq U^{ds}_{\mu\mu}\leq 1.00 
\end{equation}
for the parameter $U^{ds}_{\mu\mu}$ introduced in Eq.\ (\ref{U-def}). As expected from the discussion 
in the previous Subsection, this quantity shows a small difference from one due to the mass differences 
in Eq.~(\ref{eq:massdiff}) within the flavour-universal NP scenario. 
It will be very interesting to get much better measurements of the $B^0_d\to\mu^+\mu^-$ decay and to 
see whether they will be consistent with the picture given above.

\boldmath
\section{The Decays $B^0_s\to\tau^+\tau^-$ and $B^0_d\to\tau^+\tau^-$}\label{sec:Bstautau}
\unboldmath
\subsection{Observables}
As is evident from Eq.~(\ref{untaggedRate}), the helicity suppression of the SM rates of the
$B^0_{s,d}\to\tau^+\tau^-$ channels is essentially lifted through the large 
mass of the $\tau$ leptons. Within the SM, we obtain the following predictions:
 \begin{equation} \label{eq:BstautauSM}
\overline{\mathcal{B}}(B_s \to \tau^+\tau^-)_\text{SM} = (7.56 \pm 0.35) \times 10^{-7},
\end{equation}
\begin{equation} \label{eq:BdtautauSM}
\overline{\mathcal{B}}(B_d \to \tau^+\tau^-)_\text{SM} = (2.14 \pm 0.12) \times 10^{-8}.
\end{equation}
In order to calculate these results, we have employed the analysis of Ref.~\cite{Bsmumu-SM}, and have 
used the values of CKM and non-perturbative parameters given in Table~\ref{tab:inputs}. 

It is experimentally very challenging to reconstruct the $\tau$ leptons, in particular in the environment
of the LHC. Nevertheless, the LHCb collaboration has recently come up with the first experimental
upper limits for the corresponding branching ratios \cite{LHCb-Btautau}:
 \begin{equation} \label{eq:Bstautau-exp}
\overline{\mathcal{B}}(B_s \to \tau^+\tau^-) <  6.8\times 10^{-3} ~\mbox{(95\% C.L.)}
\end{equation}
\begin{equation} \label{eq:Bdtautau-exp}
{\mathcal{B}}(B_d \to \tau^+\tau^-) <  2.1 \times 10^{-3} ~\mbox{(95\% C.L.)}.
\end{equation}
These results are in fact the first direct constraint for $B^0_s \to \tau^+\tau^-$ and the world's best
limit for $B^0_d \to \tau^+\tau^-$.

The SM predictions for $\mathcal{A}_{\Delta\Gamma_s}^{\tau\tau}$ and $\tau^s_{\tau\tau}$ take
the same values as their $B^0_s\to\mu^+\mu^-$ counterparts:
\begin{equation}\label{Bstautau-ADG}
\mathcal{A}_{\Delta\Gamma_s}^{\tau\tau}|_{\rm SM}=+1,\quad 
\tau^s_{\tau\tau}|_{\rm SM}=\frac{\tau_{B_s}}{1-y_s} = (1.61 \pm 0.06) \, \hbox{ps}.
\end{equation}

\subsection{New Physics Benchmark Scenario}\label{sec:NP-Btautau}
Let us now have a look at the NP effects for the $B^0_{s,d}\to\tau^+\tau^-$ modes
within the benchmark scenario introduced in Subsection~\ref{ssec:BSM}. Here we obtain
the following coefficients:
\begin{equation}\label{P-expr-tau}
P^s_{\tau\tau} = \left(1-\frac{m_\mu}{m_\tau}\right){\cal C}_{10}+\frac{m_\mu}{m_\tau}P^s_{\mu\mu}
\end{equation}
\begin{equation}\label{S-expr-tau}
S^s_{\tau\tau} = \frac{m_\mu}{m_\tau}
\sqrt{\frac{1-4\frac{m_\tau^2}{M_{B_s}^2}}{1-4\frac{m_\mu^2}{M_{B_s}^2}}}S^s_{\mu\mu}. 
\end{equation}
Consequently, the NP correction to ${\cal C}_{10}$ and those proportional to $P^s_{\mu\mu}$ and
$S^s_{\mu\mu}$ are strongly suppressed through the ratio of the muon and tau masses, which is 
given as follows \cite{Mohr:2014}:
\begin{equation}
\frac{m_{\mu}}{m_{\tau}} = 0.059,
\end{equation}
and yields 
\begin{equation}\label{eq:Rtautau}
0.8\leq \overline{R}^s_{\tau\tau} \leq 1.0, 
\quad 0.995 \leq \mathcal{A}_{\Delta\Gamma_s}^{\tau\tau}\leq 1.000.
\end{equation}

The impact of NP in the $B^0_d\to\tau^+\tau^-$ decay is very similar to its $B^0_s$ counterpart, 
with $\overline{R}^d_{\tau\tau}$ taking the same values as  in Eq.~(\ref{eq:Rtautau}). 
Introducing a parameter $U^{ds}_{\tau\tau}$ in analogy to Eq.~(\ref{U-def}), we obtain 
\begin{equation}
1.000 \leq U^{ds}_{\tau\tau}\leq 1.002
\end{equation}
In view of the challenges related to the reconstruction of the $\tau$ leptons, the NP effects 
arising in the general flavour-universal NP scenario cannot be distinguished from the SM case, 
unless there is unexpected experimental progress. 

It is interesting to have a quick look at the picture in the MSSM with MFV described by 
Eqs.~(\ref{P-expr-MSSM}) and (\ref{S-expr-MSSM}). As was pointed out in Ref.~\cite{ANS}, 
the measured value of the ratio $\overline{R}^s_{\mu\mu}$ gives a twofold solutions for the 
parameter $A_q$ introduced in Eq.~(\ref{P-expr-MSSM}). We find
\begin{equation}\label{eq:As}
A_s = 0.09 \pm 0.09  \quad \lor \quad 0.98  \pm 0.09,
\end{equation}
corresponding to the observables $\mathcal{A}_{\Delta\Gamma_s}^{\mu\mu}\sim+1$ (as in the SM) 
and $\mathcal{A}_{\Delta\Gamma_s}^{\mu\mu}\sim-1$, respectively. 
These solutions give 
\begin{equation}
\overline{R}^s_{\tau\tau}|_{\rm MSSM}^{\rm MFV}=0.84\pm 0.16  \quad \lor \quad  0.47\pm 0.09, 
\end{equation}
and correspond to the branching ratios
\begin{equation}
\mathcal{B}(B_s\rightarrow \tau^+\tau^-)|_{\rm MSSM}^{\rm MFV}
=(6.4\pm 1.2)\times 10^{-7}  \quad \lor \quad  (3.5\pm 0.7)\times 10^{-7},
\end{equation}
respectively. We observe that the large $\tau$-lepton mass has a significant impact on these
quantities, in particular in the case $A_s\sim1$. 

\boldmath
\section{The Decays $B^0_s\to e^+e^-$ and $B^0_d\to e^+e^-$}\label{sec:Bsee}
\unboldmath
\subsection{Observables}
The most recent SM predictions for the $B^0_{s,d}\to e^+e^-$ decays were given in Ref.~\cite{Bsmumu-SM}.
Using the updated input parameters in Table~\ref{tab:inputs}, we obtain the following results:
\begin{equation} \label{eq:BseeSM}
\overline{\mathcal{B}}(B_s \to e^+ e^-)_\text{SM} = (8.35 \pm 0.39) \times 10^{-14},
\end{equation}
\begin{equation} \label{eq:BdeeSM}
\overline{\mathcal{B}}(B_d \to e^+ e^-)_\text{SM} = (2.39 \pm 0.14 ) \times 10^{-15}.
\end{equation}
The extremely small values of these branching ratios with respect to their $B^0_{s,d}\to\mu^+\mu^-$
counterparts arise from the helicity suppression due to the tiny electron mass, corresponding to 
an overall multiplicative factor $m^2_{e}$ in the expressions for $\mathcal{B}(B_{s,d} \to e^+ e^-)$. 
Consequently, within the SM, these decays appear to be out of reach from the experimental point of 
view, which seems to be the reason for the fact that these channels have so far essentially not played 
any role in the exploration of the quark-flavour sector. 

Concerning the experimental picture, the CDF collaboration reported the following upper bounds
($90\%$ C.L.) back in 2009 \cite{CDF-Bsee}:
\begin{equation} \label{eq:BseeSM-bound}
\overline{\mathcal{B}}(B_s \to e^+ e^-) < 2.8 \times 10^{-7},
\end{equation}
\begin{equation} \label{eq:BdeeSM-bound}
\overline{\mathcal{B}}(B_d \to e^+ e^-) < 8.3 \times 10^{-8}.
\end{equation}
Consequently, any attempt to measure the SM branching ratios for the rare decays 
$B^0_s \to e^+ e^-$ and $B^0_d \to e^+ e^-$ would require a future improvement by nearly 
six orders of magnitude. The LHC experiments have not yet reported any searches for these modes. 

\subsection{New Physics Benchmark Scenario}\label{sec:NPBsee}
Let us now consider the flavour-universal NP scenario introduced in 
Subsection~\ref{subsec:BsmumuNP}. In this framework, we obtain the following coefficients:
\begin{equation}\label{P-expr-e}
P^s_{ee} = \left(1-\frac{m_\mu}{m_e}\right){\cal C}_{10}+\frac{m_\mu}{m_e}P^s_{\mu\mu}
\end{equation}
\begin{equation}\label{S-expr-e}
S^s_{ee} = \frac{m_\mu}{m_e}
\sqrt{\frac{1-4\frac{m_\tau^2}{M_{B_s}^2}}{1-4\frac{m_\mu^2}{M_{B_s}^2}}}S^s_{\mu\mu}. 
\end{equation}
While we got a suppression of the NP effects in the $B^0_{s,d}\to\tau^+\tau^-$ decays through the
large $\tau$ mass, (see (\ref{P-expr-tau}) and (\ref{S-expr-tau})), we get now a huge enhancement 
thanks to the tiny electron mass \cite{Mohr:2014}:
\begin{equation}
\frac{m_{\mu}}{m_{e}} = 206.77,
\end{equation}
as the (pseudo)-scalar NP contributions lift the helicity suppression of the extremely small
SM branching ratio.

The enhancement with respect to the SM value is characterized by 
\begin{equation}
\overline{R}^s_{ee}= y_+\left[ {\cal C}_{10} -
\frac{m_{\mu}}{m_{e}} \left( {\cal C}_{10} - P^s_{\mu\mu} \right) \right]^2 
+ y_{-} \left(\mathcal{M}^s_{e,\mu}\right)
\left[\frac{m_{\mu}}{m_{e}}  S^s_{\mu\mu}\right]^2
\end{equation}
with
\begin{eqnarray}
y_{+}\equiv \frac{1 + y_s\cos\phi^{\rm NP}_s}{1+y_s} ,&& \quad y_{-}\equiv
\frac{1 - y_s\cos\phi^{\rm NP}_s}{1+y_s}
\end{eqnarray}
and
\begin{eqnarray}
\mathcal{M}^s_{e,\mu}\equiv\frac{1-4\frac{m^2_e}{M^2_{B_s}}}{1-4\frac{m^2_{\mu}}{M^2_{B_s}}},
\end{eqnarray}
while
\begin{equation}
\frac{\mathcal{A}_{\Delta\Gamma_s}^{ee}}{\cos\phi^{\rm NP}_s}= \frac{\left[
{\cal C}_{10} -\frac{m_{\mu}}{m_{e}} 
\left( {\cal C}_{10} - P^s_{\mu\mu} \right)
\right]^2 - \mathcal{M}^s_{e,\mu}\left[\frac{m_{\mu}}{m_{e}}  S^s_{\mu\mu}\right]^2
}
{\left[
{\cal C}_{10} -\frac{m_{\mu}}{m_{e}} 
\left( {\cal C}_{10} - P^s_{\mu\mu} \right)
\right]^2 + \mathcal{M}^s_{e,\mu}\left[\frac{m_{\mu}}{m_{e}}  S^s_{\mu\mu}\right]^2
}
.
\end{equation}

It is particularly interesting to consider the ratio between the $B^0_s\to e^+e^-$ and the 
$B^0_s\to \mu^+\mu^-$ branching ratios: 
\begin{eqnarray}\label{eq:RatBdeeBdmumu}
{\cal R}^{ee}_{s,\mu\mu}\equiv
\frac{\overline{\mathcal{B}}(B_s \rightarrow e^+e^-)}{\overline{\mathcal{B}}(B_s \rightarrow \mu^+\mu^-) }
&=& \frac{y_+\left(\mathcal{M}^s_{e,\mu}\right)^{\frac{1}{2}}\left[\frac{m_e}{m_{\mu}} {\cal C}_{10} - 
\left({\cal C}_{10} - P^s_{\mu\mu} \right) \right]^2 + y_{-} \left( \mathcal{M}^s_{e,\mu}\right)^{\frac{3}{2}} 
\left( S^s_{\mu\mu}\right)^2 }
{y_+ (P^s_{\mu\mu})^2 + y_{-}(S^s_{\mu\mu})^2 } \nonumber\\
&\approx&\frac{\left({\cal C}_{10}
-P^s_{\mu\mu}\right)^2+(S^s_{\mu\mu})^2}{(P^s_{\mu\mu})^2+(S^s_{\mu\mu})^2},
\end{eqnarray}
where we used $\cos\phi^{\rm NP}_s=1$ (see Eq.\ (\ref{cosNP})) and neglected the effects
associated with $y_s$ and the tiny mass ratio $m_e/m_{\mu}\sim 0.005$.  As the decay constants 
and CKM matrix elements cancel in ${\cal R}^{ee}_{s,\mu\mu}$, this ratio is a theoretically clean quantity. 
Moreover, its measurement at the LHC is not affected by the ratio $f_s/f_d$ of fragmentation functions 
\cite{FST}, which is an advantage from the experimental point of view. 

It is instructive to consider a situation with ${\cal C}_{10}=1$ and 
\begin{equation}
(P^s_{\mu\mu})^2+(S^s_{\mu\mu})^2\approx1,
\end{equation}
which corresponds to a $B^0_s\to\mu^+\mu^-$ branching ratio as in the SM and is consistent with
the current experimental situation. We obtain then
\begin{equation}
{\cal R}^{ee}_{s,\mu\mu}\approx 2(1-P^s_{\mu\mu}),
\end{equation}
which yields
\begin{equation}\label{range-simple}
0\lsim {\cal R}^{ee}_{s,\mu\mu} \lsim 4
\end{equation}
as $P^s_{\mu\mu}$ varies between $-1$ and $+1$ while moving on the unit circle in the 
$P^s_{\mu\mu}$--$S^s_{\mu\mu}$ plane. It is also interesting to note that ${\cal C}_{10}=0$
would result in ${\cal R}^{ee}_{s,\mu\mu}\approx 1$ independently of the values of $P^s_{\mu\mu}$
and $S^s_{\mu\mu}$ on the unit circle, 
as can be seen in Eq.\ (\ref{eq:RatBdeeBdmumu}). These simple considerations
illustrate nicely the possible spectacular enhancement of the $B^0_s\to e^+e^-$ branching ratio
with respect to the SM prediction.

\begin{figure}
\centering
\includegraphics[height=8.0cm]{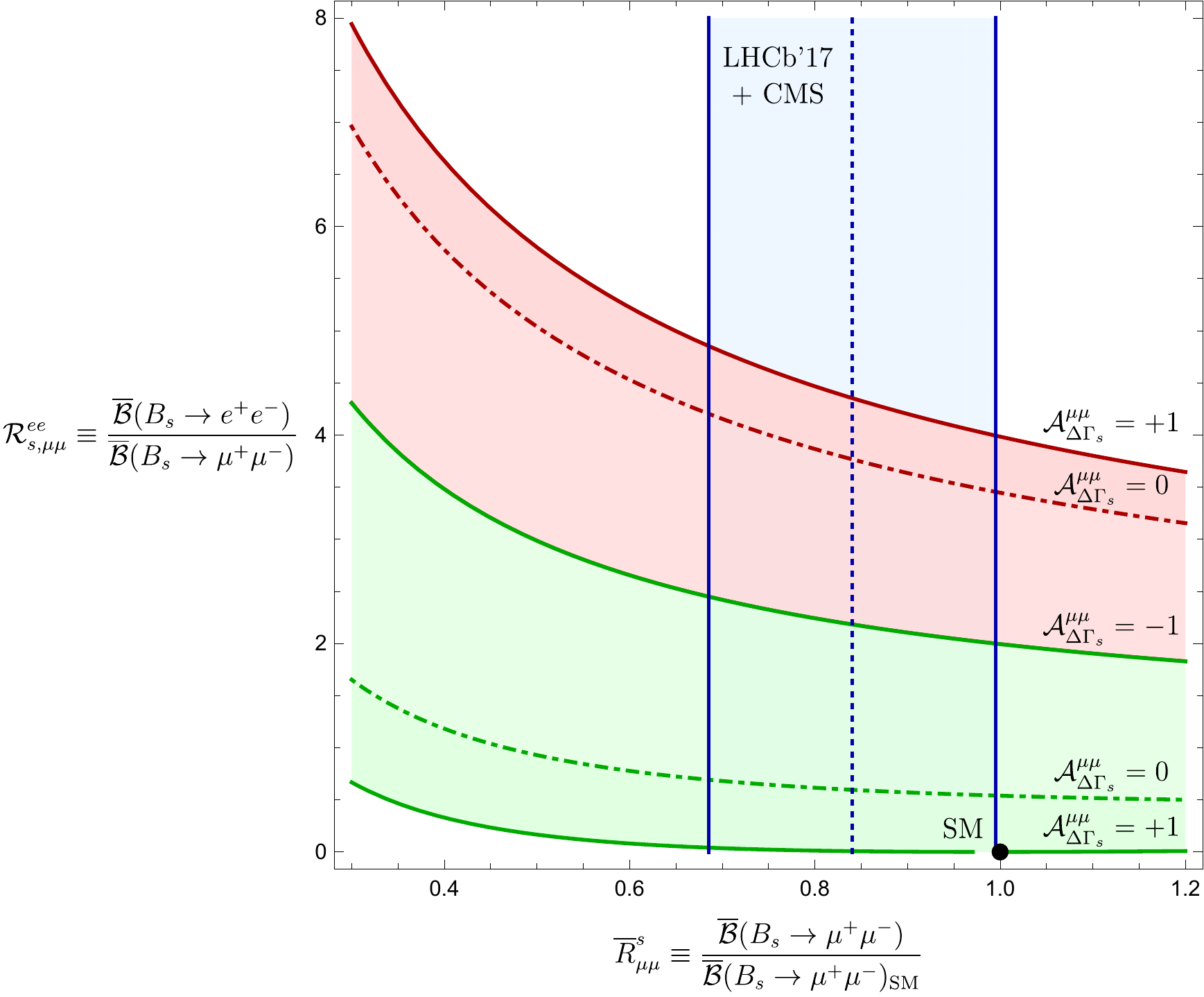}
\caption{Illustration of the allowed range for the ratio ${\cal R}^{ee}_{s,\mu\mu}$ as a function of 
$\bar{R}^s_{\mu\mu}$ in the flavour-universal NP scenario with ${\cal C}_{10}=1$. We show 
also contours for various values of 
$\mathcal{A}_{\Delta\Gamma_s}^{\mu\mu}$. As in Fig.~\ref{fig:showbrdeebrBarsmumu}, 
the green and red bands correspond to positive and negative solutions for $P_{\mu\mu}^s$, 
respectively.}\label{fig:showbrBarseebrBarsmumurat}
\end{figure}

In Fig.~\ref{fig:showbrBarseebrBarsmumurat}, we put these considerations on a more quantitative ground,
showing the allowed region for ${\cal R}^{ee}_{s,\mu\mu}$ as a function of $\overline{R}^s_{\mu\mu}$.
We give also contours for various values of $\mathcal{A}_{\Delta\Gamma_s}^{\mu\mu}$. If we vary
this observable within the range in Eq.\ (\ref{ADG-range}), we obtain
\begin{equation}\label{eq:enhacBrBsee}
0\leq {\cal R}^{ee}_{s,\mu\mu} \leq 4.8,
\end{equation}
which is consistent with (\ref{range-simple}).

In the case of the branching ratio of the $B^0_s \rightarrow e^+e^-$ channel, we obtain uncertainties 
from the decay constant $f_{B_s}$, CKM factors and the $B_s$ lifetime $\tau_{B_s}$. The range in 
Eq.~(\ref{eq:enhacBrBsee}) corresponds to 
\begin{equation}\label{Bsee-range-UNP}
0\leq \overline{\mathcal{B}}(B_s \rightarrow e^+e^-)\leq1.4\times 10^{-8}
\end{equation}
with 
\begin{equation}
0\leq \overline{R}^s_{ee}\leq 1.7\times 10^{5},
\end{equation}
showing an impressive lift of the helicity suppression with respect to the SM value in 
Eq.~(\ref{eq:BseeSM}). The observable $\mathcal{A}^{ee}_{\Delta \Gamma_s}$ is just constrained
within its general range $[-1,+1]$.

The pattern of the NP effects in $B^0_d\to e^+e^-$ is very similar to the situation in its $B^0_s$
counterpart within the considered framework, yielding
\begin{equation}\label{BR-Bd-sol-NP}
0\leq\overline{\mathcal{B}}(B_d \rightarrow e^+e^-)\leq 4.0 \times 10^{-10}.
\end{equation}

In analogy to the $B^0_s\to\tau^+\tau^-$ case, it is instructive to have also a look at the 
MSSM with MFV, which is described by Eqs.~(\ref{MSSM-rel}--\ref{S-expr-MSSM}) and 
the solutions in Eq.\ (\ref{eq:As}). In this framework, we get 
\begin{equation}
{\cal R}^{ee}_{s,\mu\mu}|_{\rm MSSM}^{\rm MFV}
\approx \Bigl(\frac{m_e}{m_{\mu}}\Bigl)^2=2.3\times 10^{-5},
\end{equation}
where we have neglected tiny $m_\mu^2/M_{B_s}^2$ and $m_e^2/M_{B_s}^2$ corrections. 
Consequently, the ratio of the branching ratios is as in the SM, and we obtain from the measured 
$B^0_s\to\mu^+\mu^-$ branching ratio:
\begin{eqnarray}
\overline{R}^s_{ee}|_{\rm MSSM}^{\rm MFV}&=&0.84\pm 0.16, \nonumber\\
\mathcal{B}(B_s\rightarrow e^+e^-)|_{\rm MSSM}^{\rm MFV}&=&(7.0 \pm 1.4)\times 10 ^{-14}.
\end{eqnarray}

It will be very interesting to search for the $B^0_{s,d}\to e^+e^-$ decays, in particular in view of the
exciting situation that the CDF upper bound from 2009 is only about a factor of 20 above 
the upper bound (\ref{Bsee-range-UNP}) in the flavour-universal NP scenario. Should the 
$B^0_{s,d}\to e^+e^-$ decays actually be observed with hugely enhanced branching ratios, 
the MSSM with MFV would be ruled out.

\boldmath
\section{Conclusions}\label{sec:concl}
\unboldmath
Leptonic rare decays of $B^0_s$ and $B^0_d$ mesons play an outstanding role for testing the SM.
The main actors have so far been the $B^0_s\to\mu^+\mu^-$ and $B^0_d\to\mu^+\mu^-$ modes,
where the former decay is now well established in the LHC data and first signals for the latter channel
were reported. Very recently, LHCb has presented the first measurement of the effective lifetime of
$B^0_s\to\mu^+\mu^-$, and upper bounds for the $B^0_s\to\tau^+\tau^-$ and $B^0_d\to\tau^+\tau^-$
modes. The experimental constraint for $\mathcal{B}(B_s\to e^+e^-)$ is six orders of magnitude
above the SM prediction, and was obtained by CDF in 2009.

We have given a state-of-the-art discussion of the interpretation of the
$B^0_{s,d}\to\mu^+\mu^-$ data. However, the main focus was on the decays with tau leptons and electrons
in the final state, addressing the question of how much space for NP effects is left by the current data,
in particular the observation of $B^0_s\to\mu^+\mu^-$. In order to explore this issue, which is in
general very involved, we have considered a NP scenario as a benchmark with flavour-universal Wilson 
coefficients of the four-fermion operators, and assumed that NP enters through (pseudo)-scalar 
contributions, which is the key domain of the $B^0_{s,d}\to\ell^+\ell^-$ decays. We may then convert 
the experimental value of the $B^0_s\to \mu^+\mu^-$ branching ratio into predictions for the other 
$B^0_{s,d}\to\ell^+\ell^-$ channels. It will be important to significantly reduce the uncertainty of the 
measurement of the observable $\mathcal{A}_{\Delta\Gamma_s}^{\mu\mu}$ in the future, 
which will have an impact on the allowed regions for these channels.  

In this scenario, we find that the NP effects are strongly suppressed by the mass ratio $m_\mu/m_\tau$
in the $B^0_{s,d}\to\tau^+\tau^-$ decays, thereby resulting in a picture which is essentially
as in the SM. On the other hand, the NP effects are amplified in the $B^0_s\to e^+e^-$ channel 
due to the mass ratio $m_\mu/m_e$. In this case, the helicity suppression is lifted by the new 
(pseudo)-scalar contributions, while the branching ratio of $B^0_s\to\mu^+\mu^-$ stays 
in the regime of the SM value, following from the current measurement of this channel. It is exciting
to find values of the $B^0_s\to e^+e^-$ branching ratio about 5 times as large as the 
$B^0_s\to\mu^+\mu^-$ branching ratio, which is a factor of about 20  below the CDF limit. The ratio 
of the $B^0_s\to e^+e^-$ and $B^0_s\to\mu^+\mu^-$ branching ratios is a theoretically 
clean quantity, having also advantages from the experimental point of view. 

Due to the helicity structure of possible NP contributions, $B^0_s\to e^+e^-$ is in general a very 
sensitive probe of physics beyond the SM with new (pseudo)-scalar contributions. As this decay 
has essentially not received any attention since the CDF analysis from 2009, it would be most 
interesting to search for $B^0_s\to e^+e^-$ in the LHC data, with the possibility of finding a signal
which would give us unambiguous evidence for New Physics. In such a situation, the MSSM with
MFV would be excluded. 

In order to get the full picture, also $B^0_s\to\tau^+\tau^-$ and the corresponding $B^0_d$ modes
should receive full attention at the LHC and the future Belle II experiments. We are excited to see
new results -- in particular searches for the $B^0_{s,d}\to e^+e^-$ decays -- which may eventually
open a window to the physics beyond the Standard Model.


\section*{Acknowledgements}
This research has been supported by the Netherlands Foundation
for Fundamental Research of Matter (FOM) programme 156, ``Higgs as Probe and Portal'', and
by the National Organisation for Scientific Research (NWO).


%
%
%

%
%
%
\end{document}